\documentclass{LMCS} 
\pdfoutput=1
\usepackage[latin1]{inputenc}
\usepackage{ae,aecompl}      

\usepackage{eucal}           
\usepackage{graphicx,color}  
\usepackage{varioref}
\usepackage{enumerate,hyperref}

\newcommand{\ie}{\textsl{i.e.}}
\newcommand{\eg}{\textsl{e.g.}}%
\newcommand{\etal}{\textsl{et al.}}%
\newcommand{\Nesetril}[0]{Ne\-\v{s}e\-t\v{r}il}
\newcommand{\integerset}[0]{\ensuremath{\mathbb{N}}}
\newcommand{\rationalset}[0]{\ensuremath{\mathbb{Q}}}
\newcommand{\realset}[0]{\ensuremath{\mathbb{R}}}

\newtheorem*{claim}{Claim}
\newtheorem*{thm*}{Theorem}

\def\doi{5 (2:13) 2009}
\lmcsheading%
{\doi}
{1--25}
{}
{}
{Jan.~\phantom{0}3, 2007}
{Jun.~\phantom{0}2, 2009}
{}   

\begin{document}
\title{Universal Structures and the logic of Forbidden Patterns.}

\author[F.~R.~Madelaine]{Florent R.~Madelaine}
\address{Univ Clermont1, EA2146 \\
  Laboratoire d'algorithmique et d'image de Clermont-Ferrand \\
  Aubière, F-63170, France}
\email{florent.madelaine@u-clermont1.fr}
\thanks{Most of this work was undertaken at Durham University and was
  partially supported by EPSRC grant EP/C54384x/1}
\titlecomment{Extended version of the conference papers~\cite{MinorClosed}
  and \cite{BoundedDegree}.}
\begin{abstract}
  We show that \emph{forbidden patterns problems}, when restricted to
  some classes of input structures, are in fact \emph{constraint satisfaction
    problems}. This contrasts with the case of unrestricted input
  structures, for which it is known that there are forbidden patterns
  problems that are not constraint satisfaction problems.
  We show that if the input comes from a class of connected
  structures with \emph{low tree-depth decomposition} then
  \emph{every} forbidden patterns problem is in fact a constraint
  satisfaction problem. In particular, our result covers input
  restrictions such as: structures of bounded degree, planar graphs,
  structures of bounded tree-width and, more generally, classes
  definable by at least one forbidden minor.
  This result can also be rephrased in terms of expressiveness of the
  logic MMSNP, introduced by Feder and Vardi in relation with
  constraint satisfaction problems. 
  Our approach follows and generalises that of \Nesetril{} and
  Ossona de Mendez's, who investigated \emph{restricted dualities},
  which corresponds in our setting to investigating the restricted
  case when the considered forbidden patterns problems are captured by a
  first-order sentence. Note also that our result holds in the general
  setting of problems over arbitrary relational structures (not just
  for graphs). 
\end{abstract}

\maketitle

\section*{Introduction}\label{sec:introduction}
Constraint satisfaction problems have been first investigated in
artificial intelligence as a generic concept covering a wide range of
combinatorial problems. The input of such a problem consists of a set
of variables, a set of values for these variables and a set of
constraints between these variables; the question is to decide whether
there is an assignment of values to the variables that satisfies all
the constraints. Note that these sets (of values, variables and
constraints) are usually assumed to be finite, thus in particular such
problems do not cover constraints over integers or over
reals. However, this framework remains general enough to cover a
variety of well-known problems such as Boolean satisfiability, graph
3-colourability,\break  \noindent and conjunctive query evaluation; and, more
importantly for us as it is the approach we shall adopt in this paper,
constraint satisfaction problems can be phrased as \emph{homomorphism
problems}. For further details, we suggest the
survey~\cite{KolaitisVardiBook05} which gives a detailed background on
the intimate connection between constraint satisfaction problems,
database theory and finite model theory. Graph homomorphisms and
related problems have received considerable attention in recent years
as a topic in combinatorics, and the monograph~\cite{HellNesetrilBook}
serves as a good survey of the area.

The theoretical investigations of constraint satisfaction problems
have been concerned mostly with computational complexity; in
particular, with the \emph{dichotomy conjecture}, that asserts that every constraint
satisfaction problem is either tractable (polynomial time
decidable) or intractable (NP-complete). This conjecture is supported
by early results in the Boolean case~\cite{Schaefer} and in the case of
graph homomorphism~\cite{HellNes}; and, by later results using tools
from universal algebra~\cite{AlgebraCSP}. So, despite the fact that constraint
satisfaction problems capture numerous well-known problems in NP,
the dichotomy conjecture suggests a fundamental difference with the
class NP (recall Ladner's Theorem~\cite{Ladner75} which states that if
P is different from NP then there is an infinite number of distinct
polynomial-time equivalence classes in NP).
 
Investigations on the fundamental properties of constraint
satisfaction problems have also focused on their descriptive
complexity. Based upon Fagin's logical characterisation of NP as those
problems expressible in the existential fragment of second order
logic, Feder and Vardi attempted to find a large 
(syntactically-defined) sub-class of NP which exhibits the dichotomy
property~\cite{FederVardi}. What emerged from Feder  and Vardi's
consideration was the logic MMSNP (short for Monotone Monadic SNP
without inequalities) defined by imposing syntactic restrictions upon
the existential fragment of second-order logic. Though syntactically
defined this logic is very combinatorial in nature, in the sense that every
sentence corresponds to a finite set of coloured obstructions: this
was made precise in~\cite{ThePaperLong}, where Madelaine and Stewart
introduced a new class of combinatorial problems, the so-called
\emph{forbidden patterns problems}, and proved that every sentence of
MMSNP defines a finite union of forbidden patterns problems (since we
only deal with decision problems in this paper, we equate a problem
with the set of its yes-instances).
Feder and Vardi were unable to prove that MMSNP has the dichotomy
property, a question which remains open, but showed that this logic is
``computationally'' equivalent to the class of constraint satisfaction
problems. This result, together with Kun's
derandomisation~\cite{GaborKun} of a 
particular graph construction used in Feder and Vardi's reduction,
implies that MMSNP has the dichotomy property if, and only, if the dichotomy
conjecture holds. So, one
could argue that MMSNP and the class of constraint satisfaction
problems are essentially the same. 

However, as was first observed by
Feder and Vardi, the logic MMSNP is too strong: there are forbidden
patterns problems which are not constraint satisfaction
problems~\cite{FederVardi,FirstPaper}. Moreover, Bodirsky and
Dalmau~\cite{BodirskyDalmauStacs06} showed that forbidden 
patterns problems are in fact examples of well-behaved constraint 
satisfaction with a countable set of values (which shall be referred as
\emph{infinite constraint satisfaction problems} thereafter). 
So, in the context of descriptive complexity, the logic MMSNP 
and constraint satisfaction problems are rather different. 
In~\cite{ThePaperLong}, Madelaine and Stewart gave an effective
characterisation of forbidden patterns problems that are constraint
satisfaction problems: given a forbidden patterns  problem, we can
decide  whether it is a finite or infinite constraint satisfaction
problem; and, in the former case, we can compute effectively a description of
this problem as a finite constraint satisfaction problem. Since the
transformation of a sentence of MMSNP into a finite union of forbidden
patterns problems  is also effective, as a corollary, we can decide
whether  a given sentence of MMSNP defines a finite union of (finite) 
constraint satisfaction problems, or defines a finite union of
infinite constraint satisfaction problems.

The question of expressivity of the logic MMSNP with respect to the
class of constraint satisfaction problems is also being studied in a
different guise in structural combinatorics. For example,
the above result, which delineates the border between forbidden patterns
problems and constraint satisfaction problems (respectively, MMSNP and
finite union of constraint satisfaction problems), subsumes the 
characterisation of duality pairs (respectively, finite dualities)
obtained by Tardif and \Nesetril{}~\cite{tardifnesetrilduality}
(see also the sequel paper~\cite{FoniokNesetrilTardif06}). 
Moreover, Kun and \Nesetril{} give a new proof of the characterisation of 
forbidden patterns problems that are constraint satisfaction
problems~\cite{KunNesetril06}. Their elegant approach involves lifting
dualities to a more complex form of dualities. However, contrary to Madelaine and
Stewart's result, note that this approach is not effective.  
The present paper is motivated by another kind of duality known as \emph{restricted
  duality}, which corresponds to restricting the studied problems to a
particular class of inputs. Contrary to the case of duality, for
particular restrictions, \textsl{e.g.} for bounded
degree~\cite{HaggkvistHell} or planar graphs~\cite{NesetrilPomStoc06}, it turns
out that we have \emph{all restricted dualities}. This provokes the following
question, which is the object of this paper. \emph{For which input
  restrictions, do \emph{all} forbidden patterns problems become constraint
  satisfaction problems}?

Before we can discuss in more detail our contribution, let us
introduce more precisely duality and restricted duality.
A \emph{duality pair} is a pair $(F,H)$ of structures such that, for
every structure $G$, there is no homomorphism from $F$ to $G$ (we say
that $G$ is $F$-mote) if, and only if, there is a homomorphism from
$G$ to $H$ (we say that $G$ is homomorphic to $H$). There are
structures $F$, for which there is no $H$ such that $(F,H)$ is a
duality pair (for example, choose $F$ to be the triangle). In this
sense, we do not have all dualities. However, when we turn to
restricted duality, that is when $G$ in the above definition ranges 
over a particular class of structures, there are examples in which we have
all dualities. For example,  H{\"a}ggkvist and
Hell~\cite{HaggkvistHell} showed how to build a finite ``universal'' graph
$H$ for the class of $F$-mote graphs of bounded degree $b$. That is, any
degree $b$ graph $G$ is $F$-mote if, and 
only if, there is a homomorphism from $G$ to $H$. Note that the usual
notion of universal graph, as used by
Fra{\"{\i}}ss{\'e}~\cite{Fraisse} is for induced substructures, not 
homomorphism, and that the universal graph is usually countable. In our
context, the word universal graph refers to a \emph{finite} graph and
is universal with respect to the existence of homomorphisms.

\begin{figure}
  \centering
  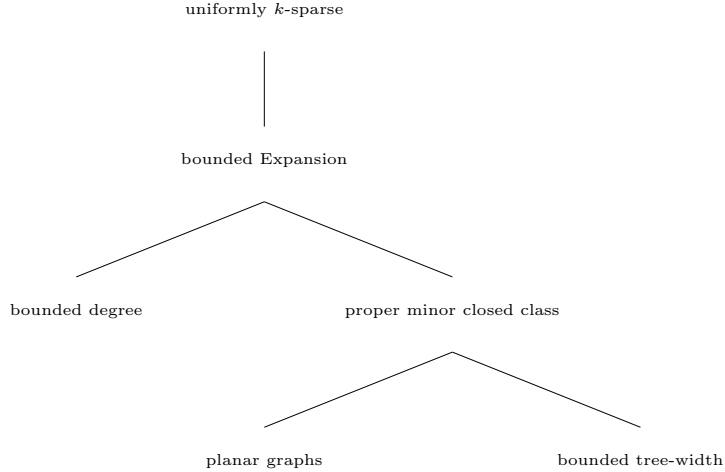
  \caption{Classes of graphs and how they compare.}
  \label{fig:comparing:classes:of:graphs}
\end{figure}

\Nesetril{} and Ossona de Mendez~\cite{ProperMinorClosedClass}
introduced the notion of \emph{tree-depth} of a graph and of \emph{low 
  tree-depth decomposition} of a class of graphs. They show a
restricted duality theorem for restriction to classes of low
tree-depth decomposition: this includes graphs of bounded degree,
planar graphs, graphs of bounded tree-width, any proper minor closed
class and, more generally, classes of bounded expansion. In 
Figure~\ref{fig:comparing:classes:of:graphs}, we have depicted how
these classes compare. See~\cite{NesetrilPomStoc06} for further
details on these classes, and for the algorithmic features that the
existence of such low tree-depth decomposition provides.  
In this paper, we generalise the approach taken
in~\cite{ProperMinorClosedClass} and show for arbitrary 
relational structures, not just for graphs, that every forbidden
patterns problem is a constraint satisfaction problem when restricted
to a class of connected structures that has a low tree-depth
decomposition. Since this case is technically much more involved than
the simple case of bounded-degree inputs, as a warm-up case to
illustrate the basic concepts, we first show that every
forbidden patterns problems restricted to connected inputs of
bounded degree becomes a constraint satisfaction problem 
(this result already appeared in a conference paper~\cite{BoundedDegree}). 
Next, we extend this result and prove that every forbidden patterns problem
when restricted to a class of connected structures with low tree-depth
decomposition is a constraint satisfaction problem. Examples of
classes of structures with this property are proper minor closed classes
(this result was originally presented in this restricted case 
in the preliminary version of this paper~\cite{MinorClosed}), classes
of bounded degree and, more generally, classes of bounded expansion. 

A well-known extension of monadic second order logic consists of
allowing monadic predicates to range over tuples of elements, rather
than just elements of a structure. This extension is denoted by
MSO$_2$ whereas  the more standard logic with monadic predicates
ranging over elements only is denoted by MSO$_1$. In general, MSO$_2$
is strictly more expressive than MSO$_1$. 
However, Courcelle~\cite{CourcelleMSO6} proved that MSO$_2$ collapses to
MSO$_1$, for some restriction of 
the input: for graphs of degree at most $k$, for graphs of tree-width
at most $k$ (for any fixed $k$), for planar graphs and, more
generally, for graphs belonging to a proper minor closed class (a
class $\mathcal{K}$ such that: a minor of a graph $G$ belongs to
$\mathcal{K}$ provided  that $G$ belongs to $\mathcal{K}$; and,
$\mathcal{K}$ does not contain all graphs). 
It is perhaps worth mentioning that
in~\cite{BoundedDegree,MinorClosed} and the 
present paper we assume a definition of forbidden patterns problems
that is more general than the original one in~\cite{ThePaperLong} in
that the new definition 
allows colourings not only of the elements of the input structure but
also of its tuples of elements. Essentially, this means that we are
now considering problems related to MMSNP$_2$, the extension of MMSNP
with ``edge'' quantification (for clarity, from now on, we denote by
MMSNP$_1$ the original logic, i.e. without edge
quantification). This means that the main result of this paper
provides us with a theorem
analogous to that of Courcelle but concerning equal expressivity of
the logics MMSNP$_1$ and MMSNP$_2$ (they both capture the class of
constraint satisfaction problems, when suitably restricted). 
Courcelle extended the above result to uniformly sparse
graphs~\cite{CourcelleMSO14}. However, as we shall see our result
fails in this case: there is a problem in MMSNP$_1$ which is not a
constraint satisfaction problem and since this problem is very
restricted, it follows in fact that uniformly $k$-sparse graphs do not
have all restricted dualities, for any fixed $k\geq 2$.

As a further motivation for the definition of MMSNP$_2$ we show that
this logic, just like MMSNP$_1$, correspond to infinite constraint satisfaction problems in the
sense of Bodirsky\cite{ManuelPhd,BodirskyDalmauStacs06}.  

The paper is organised as  follows. 
In Section~\ref{sec:preliminaries}, we define constraint satisfaction
problems, forbidden patterns problems and give some
examples. We conclude this section by a brief
review of results on restricted dualities and we state the main result
of this paper.

In Section~\ref{sec:bounded-degree}, we prove that forbidden patterns
problems are constraint satisfaction problems when restricted to
connected structures of bounded degree. This serves as a warm up case to
illustrate the main definitions and concepts.

In Section~\ref{sec:low:tree:depth:decomposition}, we prove our main
result, that is that forbidden patterns problems are constraint
satisfaction problems when restricted to a class of connected
structures that have low tree-depth decomposition. In order to do so,
we first introduce the notion of tree-depth for structures. Then, we
show that coloured structures of bounded tree-depth have bounded
cores. We conclude the proof by the construction of a finite universal
coloured structure, using the existence of a low tree-depth
decomposition for every input.

In Section~\ref{sec:logical-aspects}, we turn to logical aspects. 
We define MMSNP$_1$, recall some known results for this logic
and extend them to MMSNP$_2$.  Next, we
reformulate our main result in terms of expressivity of the logics MMSNP$_2$
and MMSNP$_1$ with respect to constraint satisfaction problems. 
We compare this result with Courcelle's result on the expressivity of
MSO$_1$ and MSO$_2$. We conclude this section by proving that  
problems in MMSNP$_2$ are also infinite constraint satisfaction
problems in the sense of Bodirsky. 

In the last section, we conclude and discuss related work and open
questions. 

\section{Preliminaries}
\label{sec:preliminaries}

\subsection{Constraint satisfaction problems and forbidden patterns problems}

Let $\sigma$ be a signature that consists of finitely many relation symbols.
From now on, unless otherwise stated, every structure considered will
be a $\sigma$-structure.  Let $S$ and $T$ be two structures. A
\emph{homomorphism} $h$ from $S$ to $T$ is a mapping from $|S|$ (the
domain of $S$)  to $|T|$ such that for every $r$-ary relation  symbol
$R$ in $\sigma$ and any elements $x_1,x_2,\ldots,x_r$ of $S$, 
if $R(x_1,x_2,\ldots,x_r)$ holds in $S$ 
then $R(h(x_1),h(x_2),\ldots,h(x_r))$ holds in $T$. If there exists a
homomorphism from $S$ to $T$ we say that $S$ is \emph{homomorphic} to $T$.
 
The \emph{constraint satisfaction problem} with \emph{template} $T$ is
the decision problem with,
\begin{enumerate}[$\bullet$]
\item input: a finite structure $S$; and,
\item question: does there exist a homomorphism from $S$ to $T$?
\end{enumerate}
We denote by CSP the class of constraint satisfaction problems
with a \emph{finite} template.
\begin{exa}
  The constraint satisfaction problem with template $K_3$ (the clique with
  three elements, \ie{} a triangle) is nothing other than the $3$-colourability
  problem from graph theory. 
\end{exa}

Let $\mathcal{V}$ (respectively, $\mathcal{E}$) be a finite set of 
\emph{vertex colours}  (respectively, \emph{edge colours}).
A \emph{coloured structure} is a triple
$(S,s^\mathcal{V},s^\mathcal{E})$, where 
$S$ is a structure, 
$s^\mathcal{V}$ is a mapping from $|S|$ to $\mathcal{V}$ and 
$s^\mathcal{E}$ is a mapping from $E(S)$ to $\mathcal{E}$ where,
$$E(S):=\bigcup_{R\in\sigma}\bigl\{(R,x_1,x_2,\ldots,x_r)
\text{ s.t. } R(x_1,x_2,\ldots,x_r)\text{ holds in } S\bigr\}.$$
Let $(S,s^\mathcal{V},s^\mathcal{E})$ and $(S',{s'}^\mathcal{V},{s'}^\mathcal{E})$
be two coloured structures. A \emph{colour-preserving homomorphism}
$h$ from $S$ to $S'$ is a homomorphism from $S$ to  $S'$  such that
${s'}^\mathcal{V}\circ h = s^\mathcal{V}$ and for every tuple
$t=(R,x_1,x_2,\ldots,x_r)$ in $E(S)$, ${s'}^\mathcal{E}(t')=
s^\mathcal{E}(t)$,  where $t':=\bigl(R,h(x_1),h(x_2),\ldots,h(x_r)\bigr)$.   
When the colours are clear from the context, we simply write that
\emph{$h$ preserves colours}. 
Note that the composition of two homomorphism that preserve colours is also a
homomorphism that preserves colours. 

A structure $S$ is \emph{connected} if it can not be partitioned into
two disjoint induced substructures. A \emph{pattern} is a finite
coloured structure $(F,f^\mathcal{V},f^\mathcal{E})$ such that $F$ is connected.
In this paper, patterns are used to model constraints in a
negative fashion and consequently, we refer to them as \emph{forbidden}
patterns. 
Let $\mathfrak{F}$ be a finite set of forbidden patterns.
We say that a coloured structure $(S,s^\mathcal{V},s^\mathcal{E})$
is \emph{valid} with respect to $\mathfrak{F}$ if, and only if, for
every forbidden pattern $(F,f^\mathcal{V},f^\mathcal{E})$ in
$\mathfrak{F}$, there does not exist any colour-preserving homomorphism $h$ 
from $F$ to $S$. \label{page:valid}

\begin{figure}[htbp]
  \centering
  \begin{tabular}{|cc|}
    \hline{}
    &\\
    \textsc{Vertex-No-Mono-Tri}&\includegraphics{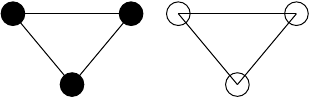}\\
    \hline{} 
    &\\
    \textsc{Tri-Free-Tri}&\includegraphics{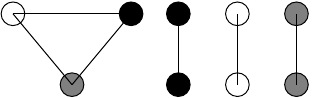}\\
    \hline{} 
    &\\
    \textsc{Edge-No-Mono-Tri}&\includegraphics{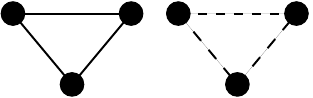}\\
    \hline 
  \end{tabular}
  \caption{Some forbidden patterns problems}
  \label{fig:forbidden:patterns}
\end{figure}

The \emph{problem with forbidden patterns $\mathfrak{F}$} is the decision
problem with,
\begin{enumerate}[$\bullet$]
\item input: a finite structure $S$
\item question: does there exist  $s^\mathcal{V}:|S|\to\mathcal{V}$
  and $s^\mathcal{E}:E(S)\to\mathcal{E}$ such that
  $(S,s^\mathcal{V},s^\mathcal{E})$ is valid with respect to $\mathfrak{F}$?
\end{enumerate}
We denote by FPP$_1$ the class of forbidden patterns problem with
vertex colours only (that is for which $\mathcal{E}$ has size one) and by FPP$_2$
the class  of forbidden patterns problems.  
\begin{exas}
  Let  $G$ be an undirected graph. It is usual to represent $G$ as a
  relational structure with a single binary relation $E$ that
  is symmetric. However, the logics considered in this paper are
  monotone and we can not express that $E$ is symmetric; therefore, we
  use a 
  different representation to encode graphs. We say that
  a structure $S$ with one binary relation symbol $E$ \emph{encodes} 
  $G$, if $|S|=V(G)$ and for any $x$ and $y$  in $V(G)$, $x$  and $y$
  are  adjacent in $G$ if, and only if, $E(x,y)$  or $E(y,x)$ holds
  in $S$. Note that this encoding is not bijective. Modulo this
  encoding, the following graph problems are forbidden patterns problems.
  \begin{enumerate}[(1)]
  \item \textsc{Vertex-No-Mono-Tri}: consists of the graphs for
    which there exists a partition of the vertex set into two sets such that
    no triangle has its three vertices occurring in a single partition.
    It was proved in~\cite{FederVardi,FirstPaper} that this problem is not in
    CSP and in~\cite{Gfree} that it is NP-complete. 
  \item \textsc{Tri-Free-Tri}: consists of the graphs that are both three
    colourable (tripartite) and in which there is no triangle. It was proved
    in~\cite{FirstPaper} that this problem is not in CSP. 
    It follows from~\cite{GaborKun} that this problem is
    NP-complete.
  \item 
    \textsc{Edge-No-Mono-Tri}: consists of the graphs for
    which there exists a partition of the edge set in two sets such that
    no triangle has its three edges occurring in a single partition.
    It is known to be NP-complete (see~\cite{GarreyJohnson}).
  \end{enumerate}
  The above examples can be formulated as Forbidden Patterns Problems. The
  corresponding sets of forbidden patterns are depicted on
  Figure~\ref{fig:forbidden:patterns}. In the case of
  \textsc{Edge-No-Mono-Tri}, the two type of colours for edges are
  depicted with dashed and full line respectively. 
\end{exas}

\subsection{Restricted Dualities}
\label{sec:bounded:degree:restricted:duality}
Let $\mathcal{C}$ be a class of structures.
We say that $\mathcal{C}$ has \emph{all restricted dualities} if, and only
if, for every finite set of connected structures $\mathfrak{F}$ there
exists a \emph{finite} structure $U$, the so-called \emph{universal
  structure}, such that for every structure $A$ in $\mathcal{C}$ there
is no homomorphism from any $F$ in $\mathfrak{F}$ to $A$ if, and only if,
$A$ is homomorphic to $U$.

The first example of a restricted duality theorem is due to H\"{a}ggvist
and Hell.
\begin{thm}{\cite{HaggkvistHell}} 
  \label{thm:restricted:duality:bounded:degree}
  Let $b$ be an integer and
  $\mathcal{C}$ be a class of graphs. If every graph in $\mathcal{C}$
  has bounded degree $b$ then $\mathcal{C}$ has all restricted dualities.
\end{thm}
More recently, \Nesetril{} and Ossona de Mendez gave a duality theorem
for proper minor closed classes.
\begin{thm}{\cite{ProperMinorClosedClass}}
  \label{thm:restricted:duality:pmcc}
  Let $M$ be a graph and
  $\mathcal{C}$ be a class of graphs. If no graph in $\mathcal{C}$
  admits $M$ as a minor then $\mathcal{C}$ has all restricted
  dualities.
\end{thm}
One of the key notions in the proof is that of a \emph{low tree-depth
decomposition} (see Section~\ref{sec:decompositions} for a
definition). More recently, the same authors have introduced the 
notion of classes of graphs of bounded expansion, 
which encompasses both classes of graphs of bounded degree and proper
minor closed classes. 
A class of graphs $\mathcal{C}$ has \emph{bounded expansion} if there
exists a function $f:\integerset \rightarrow \realset$ such that for
every graph $G$ in $\mathcal{C}$ and every $r>0$, $\nabla_r (G)$, the
so-called \emph{grad of $G$ of rank $r$} is bounded by $f(r)$, where
$\nabla_r (G) = \max \frac{|E(G|\mathcal{P})|}{|\mathcal{P}|}$.
Here,  $\mathcal{P}$ is a set of disjoint sets of vertices of $G$, each
of which induce a connected subgraph of $G$; and, $E(G|\mathcal{P})$ 
denotes the edge set of the minor of $G$ constructed by identifying the
vertices inside each set into a single vertex and deleting other
vertices and edges. These authors proved that classes of graphs of
bounded expansions have also low tree-depth decomposition and proved
the following general result.
\begin{thm}{\cite{NesetrildeMendezBoundedExpansion3}}
  \label{thm:restricted:duality:bounded:expansion}
  Let $\mathcal{C}$ be a
  class of graphs. If $\mathcal{C}$ has bounded expansion then
  $\mathcal{C}$ has all restricted dualities.
\end{thm}

\begin{exa}
  We can use restricted duality results, such as those presented above,
  and \emph{ad hoc} techniques from~\cite{BoundedDegree} to show that
  two of our examples are constraint satisfaction problems when restricted to a suitable class
  $\mathcal{K}$ (\eg{} class of graphs of bounded degree, proper minor
  closed class or, more generally, class of bounded expansion; in fact,
  any class for which we have a suitable restricted duality result).
  Let $U$ be the universal graph for $\mathfrak{F}=\{K_3\}$.
  Let $U'$ be the product of $U$ with $K_3$. Recall that this is the
  graph with edge set $|U|\times|K_3|$ and with an edge
  $E((u,x),(v,y))$ if, and only if both $E(u,v)$ in $U$ and $E(x,y)$
  in $K_3$. It is not difficult to
  check that \textsc{Tri-Free-Tri}, restricted to $\mathcal{K}$ is the
  constraint satisfaction problem with template $U'$. Similarly
  \textsc{Vertex-No-Mono-Tri} is the constraint satisfaction problem
  with template $U''$, where $U''$ is the graph that consists of two 
  copies of $U$, such that every pair of elements from different copies are
  adjacent. Note that, technically, our problems being defined over
  structures with a single binary relation $E$, we have not really
  expressed them as a constraint satisfaction problem. However,
  according to our encoding, replacing any two adjacent vertices $x$
  and $y$ in the above graphs by two arcs $E(x,y)$ and $E(y,x)$
  provides us with a suitable template. 
\end{exa}

\subsection{Restricted Coloured Dualities}
We say that $\mathcal{C}$ has \emph{all restricted coloured dualities}
if, and only if, for every finite set of forbidden patterns
$\mathfrak{F}$ there exists a 
\emph{finite} structure $U$, the so-called \emph{universal structure},
such that for every structure $A$ in $\mathcal{C}$, $A$ is valid
with respect to $\mathfrak{F}$ \footnotemark{} if, and only if, $A$ is homomorphic to
$U$. 

\footnotetext{That is there is no suitable colouring of $A$ (see
  definition of validity~\vpageref{page:valid}).}

We say that a structure $S$ has \textrm{bounded degree} $b$ if every
element of $S$ occurs in at most $b$ distinct tuples.  
We extended the previous result to restricted coloured dualities (we
provide the proof of this result for completeness in
Section~\ref{sec:bounded-degree}).
\begin{thm}{\cite{BoundedDegree}}
  \label{thm:coloured:restricted:duality:bounded:degree}
  Let $b$ be an integer and $\mathcal{C}$ be a class of
  structures. If every structure $S$ in $\mathcal{C}$ has bounded
  degree $b$ then $\mathcal{C}$ has all restricted coloured dualities.
\end{thm}

In the preliminary version of this paper~\cite{MinorClosed}, we extended
Theorem~\ref{thm:restricted:duality:pmcc} to restricted coloured
dualities. We also extended the result to structures, rather than just
graphs. Let us make precise what we understand by a proper minor
closed class of structures. Let $S$ be a structure. The \emph{Gaifman
  graph} of $S$, which we denote by $\mathcal{G}_S$ is the
\emph{loopless} graph with vertices 
$|S|$ in which two distincts elements $x$ and $y$ of $S$ are adjacent
if, and only if, there exists a tuple in a relation of $S$ in which
they occur simultaneously. Given a class $\mathcal{K}$ of structures,
we denote by $\mathcal{G}_\mathcal{K}$ the class of their Gaifman
graphs.  A class of graphs $\mathcal{G}$ is said to be a \emph{proper minor
closed class} if the following holds: first, for any graph $G$ and any
minor $H$ of $G$, if $G$ belongs to $\mathcal{G}$ then so does $H$;
and, secondly $\mathcal{G}$ does not contain all
graphs. Alternatively, $\mathcal{G}$ is proper minor closed if it
excludes at least one fixed graph $M$ as a minor, a so-called
\emph{forbidden minor}. We say that a class of structures
$\mathcal{K}$ is a \emph{proper minor closed class} if, and only if,
$\mathcal{G}_\mathcal{K}$ is a proper minor closed class.
\begin{thm}{\cite{MinorClosed}}
  \label{thm:coloured:restricted:duality:pmcc}
  Let $\mathcal{C}$ be a class of structures. 
  If $\mathcal{C}$ is a proper minor closed class then $\mathcal{C}$ has all
  restricted coloured dualities. 
\end{thm}
The key property we use in the proof of this result, namely that such
a proper minor closed class has low tree-depth decomposition, holds
also in the general case of a class of (structures) of bounded
expansion defined similarly as follows. A class of structures
$\mathcal{K}$ is of \emph{bounded expansion} if, and only if,
$\mathcal{G}_\mathcal{K}$ is of bounded expansion.
Thus, a similar proof provides us in fact with the following more general
result which subsumes also
Theorem~\ref{thm:coloured:restricted:duality:bounded:degree}.
\begin{thm}
  \label{thm:coloured:restricted:duality:bounded:expansion}
  Let $\mathcal{C}$ be a class of structures. 
  If $\mathcal{C}$ has low tree-depth decomposition (\eg{} bounded
  degree, proper minor closed class, structure of bounded expansion)
  then $\mathcal{C}$ has all restricted coloured dualities. 
\end{thm}
Section~\ref{sec:low:tree:depth:decomposition} is devoted to the proof
of this result. 
\begin{exa}
  The last of our examples \textsc{Edge-No-Mono-Tri} is also a
  constraint satisfaction problem when
  restricted to a class $\mathcal{K}$ that has low tree-depth
  decomposition by the above result.
\end{exa}

\section{Bounded Degree}
\label{sec:bounded-degree}
In this section, we give a proof of
Theorem~\ref{thm:coloured:restricted:duality:bounded:degree}. To
simplify the notation we will consider graphs only but the 
proof extends to relational structures without major difficulty.
We shall first explain the intuition and ideas behind the proof by
giving an informal outline. Let us denote the largest diameter of a
forbidden pattern by $m$. This parameter, together with the
degree-bound $b$, are constants, i.e.  they are not part of
the input of the forbidden patterns problem.  The universal graph $U$ has to contain all
possible {\it small graphs} that are {\it yes-instances of the forbidden patterns problem},
``small'' meaning ``of diameter $m + 1$'' here and thereafter. The
intuition then is that each graph $G$, which is a yes-instance of the
forbidden patterns problem, however big, can be homomorphically mapped to $U$ as such a
mapping should be only locally consistent. Given that the degree of
$G$ is bounded by $b$, we need to distinguish among no more than
$X$ vertices in a small neighbourhood, where $X$ depends on $b$ and
$m$. So we can use $X$ many different labels in constructing the
vertex set of $U$.  On the other hand, in order to define the
adjacency relation of the universal 
graph, \ie{} to correctly ``glue'' all the possible small
neighbourhoods, any vertex of $U$ should carry information not only
about its label, but also about its neighbourhood. In other words, any
such vertex will represent a small graph together with a vertex which
is the ``centre'' (or the root) of the small graph.  Thus the vertex
set of the universal graph will consists of all such rooted small
graphs, vertex- and edge-coloured in all possible ways, that are
yes-instances of the forbidden patterns problem.  Two vertices will be adjacent in $U$ if,
and only if, the graphs they represent ``agree'', i.e. have most of
their vertices with the same labels and colours and the induced
subgraphs of these vertices coincide including the edge colours; for
the precise definition of what ``agree'' means, one should see the
formal proof below.  It is now intuitively clear why a yes-instance
$G$ of the forbidden patterns problem should be homomorphic to the universal graph $U$: the
vertices of $G$ can be labelled so that any two adjacent vertices get
different labels, then one can choose a good vertex- and
edge-colouring of $G$, and because of the construction of $U$ now
every vertex $u$ in $G$ can be mapped to the vertex of $U$ that
represents the small neighbourhood of $u$ rooted at $u$. It is
straightforward to see that the mapping preserves edges.

The universal graph $U$ has a very useful property, namely every small
neighbourhood of $U$ rooted at a vertex $v$ is homomorphic to the
small graph represented by the vertex $u$ (see Lemma~\ref{lem:univ:colours}
below). This property immediately implies that a no-instance 
of the forbidden patterns problem cannot be homomorphic to $U$. Indeed, suppose for the sake of
contradiction, there is a no-instance $G$ that is homomorphic to $U$. Fix the
colouring induced by the homomorphism and observe that there is a
homomorphism from the forbidden pattern $F$ into $G$. The composition
of the two homomorphisms gives a homomorphism from $F$ into $U$, and
by the property above, by another composition, we get a homomorphism
from the forbidden pattern $F$ to 
some small graph represented by a vertex of $U$. This gives a contradiction
with our construction as we have taken small no-instances only to be
represented by the vertices of the universal graph. 
\begin{figure}[htbp]
  \centering
  \begin{tabular}{|c|c|}
    \hline{}
    &\\
    (Edge) coloured graph $S_1$
    &
    (Edge) coloured graph $S_2$\\
    \hline{} 
    &\\
    \includegraphics{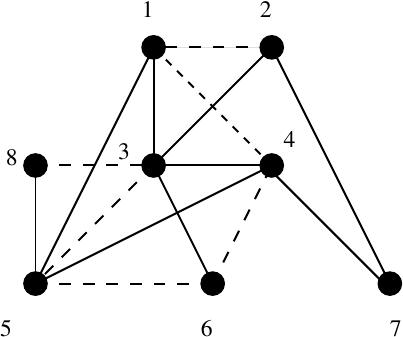} &\includegraphics{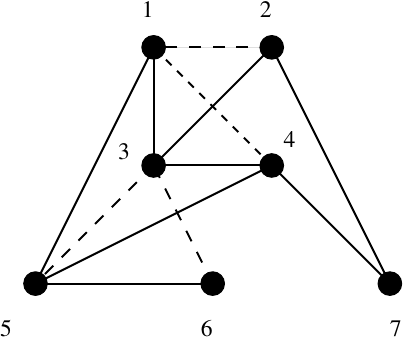}\\
    \hline \hline{} 
    &\\
    \multicolumn{1}{|p{5cm}|}{The coloured graph induced by the vertices at
      distance at most one from the vertex $1$ is identical in both $S_1$ and
      $S_2$} 
   &
    \multicolumn{1}{|p{5cm}|}{The coloured graph induced by the vertices at
      distance at most one from the vertex $2$ is identical in both $S_1$ and
      $S_2$} 
    \\
    &\\
    \hline{} 
    &\\
    \includegraphics{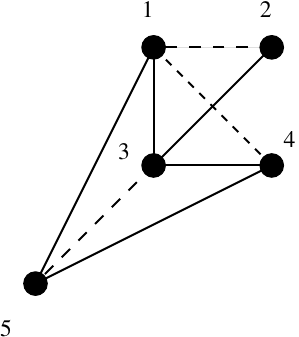} &\includegraphics{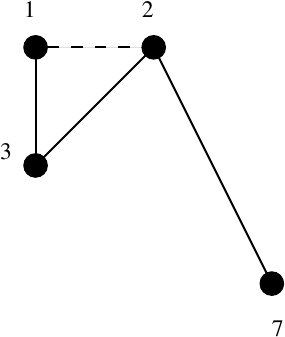}\\
    \hline \hline 
    \multicolumn{2}{|c|}{}\\ 
    \multicolumn{2}{|c|}{The template $U$ has the edge}\\
    \multicolumn{2}{|c|}{}\\ 
    \multicolumn{2}{|c|}{$(1,S_1)$\includegraphics{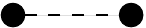}$(2,S_2)$}\\
    \multicolumn{2}{|c|}{}\\ 
    \hline 
  \end{tabular}
  \caption{Illustration of the construction for \textsc{Edge-No-Mono-Tri}.} 
  \label{fig:idea:of:proof}
\end{figure}

\begin{exa}
  We illustrate the construction of
  Theorem~\ref{thm:coloured:restricted:duality:bounded:degree} for
  our 
  final example \textsc{Edge-No-Mono-Tri} in
  Figure~\ref{fig:idea:of:proof} for inputs of 
  bounded degree $b\geq 3$. Any input $G$ can be labelled by elements from
  $\{1,2,\ldots,10\}$ such that, for every vertex $x$ in $V(G)$, the
  vertices at distance at most $2$ (the diameter of a triangle plus one) of $x$
  have a different label.    
\end{exa}

The remainder of this section is devoted to the formal proof of
Theorem~\ref{thm:coloured:restricted:duality:bounded:degree}.

Let $b$ be a positive integer and $\mathfrak{F}$ be a set of forbidden
patterns. We write $\delta(G)$ to denote the \emph{diameter} 
of a graph $G$: that is, the maximum for all vertices $x$ in $V(G)$ of the
maximum distance from $x$ to any other vertex $y$ in $V(G)$.\\
Let $m:=\max \{\delta(F) \text{ such that } (F,f^\mathcal{V},f^\mathcal{E}) \in
\mathfrak{F}\}$. Let $X:= 1 + \Sigma_{j=0}^{m} b(b-1)^j$. 

\subsection*{Construction of $U$.}
Let $\mathcal{S}$ be the set of connected graphs
$(S,s^\mathcal{V},s^\mathcal{E})$ that are valid w.r.t. $\mathfrak{F}$, such
that $V(S)$ is a subset of $\{1,2, \ldots, X\}$.
Let $U$ be the graph with: 
\begin{enumerate}[$\bullet$]
\item vertices $(v,(S,s^\mathcal{V},s^\mathcal{E}))$, where
  $(S,s^\mathcal{V},s^\mathcal{E})\in \mathcal{S}$ and $v \in S$; and, 
\item such that $(v,(S,s^\mathcal{V},s^\mathcal{E}))$ is adjacent to
  $(v',(S',{s'}^\mathcal{V},{s'}^\mathcal{E}))$ if, and only if, the 
  following holds:
  \begin{enumerate}[$-$]
  \item[$(i)$] $(v,v')$ belongs to both $E(S)$ and $E(S')$; and,
  \item[$(ii)$] the induced coloured subgraph of $S$ induced by every vertex
    at distance at most $m$ from $v$ (respectively, $v'$) is \emph{identical}
    to the induced coloured subgraph of $S'$ induced by every vertex at
    distance at most $m$ from $v$ (respectively, $v'$).
  \end{enumerate}
\end{enumerate}

We shall prove shortly that $U$ is a yes-instance witnessed by the following ``\emph{universal
  colours}'' defined as follows,
\begin{enumerate}[$\bullet$]
\item $u^\mathcal{V}(v,(S,s^\mathcal{V},s^\mathcal{E})):= s^\mathcal{V}(v)$,
  for every vertex $(S,s^\mathcal{V},s^\mathcal{E})$ in $V(U)$; and, 
\item $u^\mathcal{E}((v,(S,s^\mathcal{V},s^\mathcal{E}))(v',(S',{s'}^\mathcal{V},{s'}^\mathcal{E}))):=
  s^\mathcal{E}(v,v')$, for every edge \\
  $((v,(S,s^\mathcal{V},s^\mathcal{E}))(v',(S',{s'}^\mathcal{V},{s'}^\mathcal{E})))$
  in $E(U)$. 
\end{enumerate}
We first need the following lemma.
\begin{lem}\label{lem:univ:colours}
  Let $(v,(S,s^\mathcal{V},s^\mathcal{E}))$ in $V(U)$.
  Let $(B,b^\mathcal{V},b^\mathcal{E})$ be the coloured graph induced by all
  vertices at distance at most $m$ from
  $(v,(S,s^\mathcal{V},s^\mathcal{E}))$. 
  Then, there exists a homomorphism from $(B,b^\mathcal{V},b^\mathcal{E})$ to 
  $(S,s^\mathcal{V},s^\mathcal{E})$ that preserves the colours.
\end{lem}
\begin{proof}
  Let $(v',(S',{s'}^\mathcal{V},{s'}^\mathcal{E}))$ in $V(B)$.
  We set $h(v',(S',{s'}^\mathcal{V},{s'}^\mathcal{E})):=v'$.
  By induction on the distance from $(v,(S,s^\mathcal{V},s^\mathcal{E}))$
  to $(v',(S',{s'}^\mathcal{V},{s'}^\mathcal{E}))$, using $(i)$, it
  follows that the vertex $v'$ belongs to $V(S)$ and that $h$ is a homomorphism.
  Similarly, using $(ii)$ it follows that $h$ preserves colours.
\end{proof}
\begin{claim}
  $(U,u^\mathcal{V},u^\mathcal{E})$ is valid with
  respect to the forbidden patterns $\mathfrak{F}$.
\end{claim} 
\begin{proof}
   Assume for contradiction that
   $(U,u^\mathcal{V},u^\mathcal{E})$ is not valid and that there exists some
   forbidden pattern $(F,f^\mathcal{V},f^\mathcal{E})$ in $\mathfrak{F}$ and some
   homomorphism $f'$ from $F$ to $U$ that preserves these colours. 
   Since the diameter of $F$ is at most $m$, there exists a vertex $x$ in $V(F)$
   such that every vertex of $F$ is at distance at most $m$. 
   Hence, it is also the case for $f'(x)$ in the homomorphic image of $F$ via
   $f'$. By lemma~\ref{lem:univ:colours}, there exists a homomorphism $h$ from $F$
   to $S$, where $f'(F)=(v,(S,s^\mathcal{V},s^\mathcal{E}))$. Both homomorphisms
   are colour preserving, thus composing these two homomorphisms, we get that
   $h\circ f'$ is a colour-preserving homomorphism from $F$ to $S$.
   However, by definition of $\mathcal{S}$, the coloured graph
   $(S,{s}^\mathcal{V},{s}^\mathcal{E})$ is valid with respect to the
   forbidden patterns $\mathfrak{F}$. We reach a contradiction and the result
   follows. 
 \end{proof}
\begin{claim}
  Yes-Instances Are Homomorphic to $U$.
\end{claim}
\begin{proof}
  Let $G$ be a graph of bounded degree $b$ for which there exist 
  $g^\mathcal{V}:V(G)\to \mathcal{V}$ and 
  $g^\mathcal{E}:E(G)\to \mathcal{E}$
  such that $(G,g^\mathcal{V},g^\mathcal{E})$ is valid w.r.t. $\mathfrak{F}$.
  Since $G$ has bounded degree $b$, for every vertex $x$ in
  $V(G)$, there are at most $X-1$ vertices $y$ at distance at most $m$ from $x$.
  Therefore, there exists a map $\chi$ from $V(G)$ to $\{1,2,\ldots,X\}$ such
  that every two distinct vertices within distance $m$ or less take a different
  colour via $\chi$. 
  Thus, for every vertex $x$ in $G$, the subgraph of $G$ induced by the
  vertices at distance at most $m$ of $x$ can be identified (via the labelling
  $\chi$) to a graph $S_x$ with domain $\{1,2,\ldots,X\}$. Similarly, the
  restriction of $g^\mathcal{V} $ and $g^\mathcal{E}$ to this subgraph induce
  colour maps $s_x^\mathcal{V}$ and $s_x^\mathcal{E}$ of $S$.
  We set $a(x):=(\chi(x),(S_x,s_x^\mathcal{V},s_x^\mathcal{E}))$.
  It follows directly from the definition of $U$ that $a$ is homomorphism that
  preserve colours.  
\end{proof}
\begin{claim}
  No-instances are not homomorphic to $U$.  
\end{claim}
\begin{proof}
  Let $G$ be a graph of bounded degree $b$ that is a no instance of
  the forbidden patterns problem represented by $\mathfrak{F}$ and assume for
  contradiction that $a$ is a homomorphism from $G$ to $U$. The
  homomorphism $a$ together with the universal colouring
  $(U,u^\mathcal{V},u^\mathcal{E})$ induces colourings 
  $g^\mathcal{V}$ and $g^\mathcal{E}$ as follows:
  For every vertex $x$ in $V(G)$, set $g^\mathcal{V}(x):= u^\mathcal{V}(a(x))$;
  and, for every edge $(x,y)$ in $E(G)$, set $g^\mathcal{E}(x,y):=
  u^\mathcal{E}(a(x),a(y))$. 
  Since $G$ is a no instance, there exists a forbidden pattern
  $(F,f^\mathcal{V},f^\mathcal{E})$ in $\mathfrak{F}$ and a homomorphism $f'$
  from $F$ to $G$ that preserve these colours.
  Composing the two homomorphisms, we get that $a\circ f'$ is a homomorphism from
  $F$ to $U$ that preserves the colours of $F$ and $U$.
  This contradicts the fact that $(U,u^\mathcal{V},u^\mathcal{E})$ is valid
  w.r.t. $\mathfrak{F}$. 
\end{proof}
This concludes the proof of
Theorem~\ref{thm:coloured:restricted:duality:bounded:degree}.

\section{Low Tree-depth decomposition.}
\label{sec:low:tree:depth:decomposition}
In this section we give a proof of
Theorem~\ref{thm:coloured:restricted:duality:bounded:expansion}. 
Before moving to the formal proof, let us give an informal outline. 
Given a forbidden patterns problem, we need to build a universal
structure $U$ such that, for every input structure $S$ (that comes
from a class 
$\mathcal{K}$ that has low tree-depth decomposition), $S$ is a
yes-instance of the given forbidden patterns problem if, and only if,
$S$ is homomorphic to $U$. Recall that we use the word \emph{universal
structure} with a different meaning to that of Fra{\"{\i}}ss{\'e}, in
particular $U$ has to be finite rather than infinite and is universal
w.r.t. the existence of homomorphisms rather than induced substructures
(\textsl{i.e.} existence of embeddings). Assume for now that the
forbidden patterns problem in question has a single colour. One of the key
properties for building such a finite $U$ is that of \emph{bounded
  tree-depth}. It turns out that though the size of a structure of
bounded tree-depth may be arbitrarily large, the size of its
\emph{core} is bounded. Recall that the core of a structure $S$ is the
smallest structure that is homomorphically equivalent to $S$. Let
$Y_p$ be the disjoint union of all cores of 
structures of tree-depth at most $p$ that are valid (w.r.t. our fixed
forbidden patterns  problem). Note that $Y_p$ is finite and for any structure
$S$ of tree-depth at most $p$, $S$ is homomorphic to $Y_p$ if, and
only if, $S$ is valid. 
The next key ingredient is that the inputs have low tree-depth
decomposition: that is any input structure can be decomposed into a
fixed number of parts, say $q$, so that any $p\leq q$ parts induce a
structure of tree-depth at most $p$ (here, $q$ depends on $p$ and the
considered class of structures). So, given some input $S$ together
with such a decomposition, if the largest forbidden 
pattern has size $p$ then it suffices to check for any choice of $p$ parts
of the input, that the structure induced by these $p$-parts is valid, or
equivalently, that it is homomorphic to $Y_p$. 
Finally, we use the key concept of $p$th \emph{truncated
  product}. This concept allows us to translate the existence of
homomorphisms to a structure $T$, for any $p-1$ parts of a $p$
partitioned input $S$, to the existence of a homomorphism to the $p$th
truncated product of $T$. Hence, by taking a sequence of suitable
truncated products of $Y_p$, we get the desired finite universal
structure $U$. Note that we assumed that the forbidden patterns
problem had a single colour. In order to get our result in general, we
adapt the above ideas and concepts to coloured structures in the same
spirit as in the previous section. 

\subsection{Tree-depth and Elimination Tree of a Structure.}
Following the theory of tree-depth of graphs introduced
in~\cite{ProperMinorClosedClass}, we develop elements of a theory of
tree-depth for structures. 

Let $S$ be a structure. We denote by
$\mathcal{H}_S$ the \emph{hypergraph induced by $S$}, that has the same domain
as $S$ and whose hyperedges are the sets that consists of the elements that
occur in the tuples of the relations of $S$. If $\mathcal{H}$ is an
hypergraph and $r$ is an element of the domain of $\mathcal{H}$ then
we denote by $\mathcal{H}\setminus \{r\}$ the hypergraph obtained from
$\mathcal{H}$ by deleting $r$ from the domain of $\mathcal{H}$ and
removing $r$ from every hyperedge in  which it occurs (\eg{}
$\{a,b,r\}$ is replaced by $\{a,b\}$). A connected component
$\mathcal{H}_i$ of $\mathcal{H}_S\setminus \{r\}$ induces a
substructure $S_i$ of $S$ in a natural way: $S_i$ is the induced
substructure of $S$ with the same domain as $\mathcal{H}_i$.
 If $S$ is connected, then we say that a rooted tree $(r,Y)$ is an
\emph{elimination tree} for $S$ if, and only if, either $|S|=\{r\}$ and
$|Y|=r$, or for every component $S_i$ of $S$ ($1\leq i \leq p$) induced by
the connected components $\mathcal{H}_i$ of $\mathcal{H}_{S}\setminus
\{r\}$, $Y$ is the tree with root $r$ adjacent to subtrees
$(r_i,Y_i)$, where $(r_i,Y_i)$ is an elimination tree of $S_i$.
Let $F$ be a rooted forest (disjoint union of rooted trees).
We define the \emph{closure of $F$} $\mathrm{clos}(F,\sigma)$ to be
the $\sigma$-structure with domain $|F|$ and all tuples
$R_i(x_1,x_2,\ldots,x_{r_i})$ such that the elements mentioned 
in this tuple $\{x_i| 1\leq i \leq r_i\}$ form a chain
w.r.t. $\leq_F$, where $\leq_F$ is the 
partial order induced by $F$, \ie{} $x\leq_F y$ if, and only if, $x$
is an ancestor of $y$ in $F$.
The \emph{tree-depth} of $S$, denoted by $\mathrm{td}(S)$, is the
minimum height of a rooted forest $F$ such that $S$ is a substructure
of the closure of $F$, $\mathrm{clos}(F,\sigma)$.

These notions are closely related.
\begin{lem}\label{lem:elimination:tree:depth}
  Let $S$ be a connected structure.
  A rooted tree $(r,Y)$ is an elimination tree for $S$ if, and only if,
  $S$ is a substructure of $\mathrm{clos}(Y,\sigma)$.
  Consequently,  the tree-depth of $S$ is  the minimum height of an
  elimination tree.
\end{lem}
\proof
  We prove this result by induction on $|S|$.
  If $S$ has a single vertex $r$  then the result holds trivially.
  Assume that the above equivalence holds for every connected structure of size
  at most $n-1$ and assume that $S$ has size $n$. 

  Let $(r,Y)$ be a tree that consists of a root $r$ adjacent  to
  rooted subtrees $(r_i,Y_i)$.
  The tree $(r,Y)$ is an elimination tree for $S$ if, and only if,
  the subtrees $(r_i,Y_i)$ are elimination trees for the
  components $S_i$ induced by the connected component $\mathcal{H}_i$
  of $\mathcal{H}_S\setminus \{r\}$. By induction $S_i$ is a
  substructure of $\mathrm{clos}(Y_i,\sigma)$. Moreover every tuple
  $t$ in a relation of $S$ either does not mention $r$ and occurs in a
  single component $S_i$ or $t$ mentions $r$ and apart from $r$
  only elements from a single structure $S_i$. Hence, equivalently we
  have that $S$ is a substructure of $\mathrm{clos}(Y,\sigma)$.  
\qed

We will need the following lemma later.
\begin{lem}
  \label{lemma:gaifmangraph:structure}
  A rooted tree $(r,Y)$ is an elimination tree for a structure $S$ if,
  and only if, it is an elimination-tree for its Gaifman graph 
  $\mathcal{G}_S$.
\end{lem}
\proof
  The forward implication holds since every edge between two elements
  $x$ and $y$ in the Gaifman graph is induced by at least one tuple
  in some relation of $S$ in which both $x$ and $y$ occurs. Thus,  in
  particular $x$ and $y$  occur on the same branch in an elimination
  tree of $S$.

  Conversely, every tuple induces a clique in the Gaifman graph and
  all elements of a clique must occur on the same branch of an
  elimination tree.  Thus, an elimination tree for $\mathcal{G}_S$ is
  also an elimination tree for $S$.
\qed
\subsection{Tree-depth and Cores.}
We show that a coloured structure $\mathcal{K}$ of bounded tree-depth 
has a core of bounded size. Recall that a \emph{retract} of a structure
$S$ is an induced substructure $S'$ of $S$ for which there exists a
homomorphism from $S$ to $S'$. A minimal retract of a structure $S$ is
called a \emph{core} of $S$. Since it is unique up to isomorphism, we
may speak of \emph{the} core of a structure~\cite{HellNesetrilBook}.
This notion extends naturally to coloured structures~\cite{ThePaperLong}.

We say that a (colour-preserving) automorphism $\mu$ of a (coloured
structure) $S$ has the \emph{fixed-point property} if, for every
connected substructure $T$ of $S$, either $\mu(T)\cap T=\emptyset$ or
there exist an element $x$ in $T$ such that $\mu(x)=x$. We say that
$\mu$ is \emph{involuting} if $\mu\circ \mu$ is the identity.  
\begin{thm}\label{theor:non-trivial:involution}
  There exists a function $\eta :\integerset\times \integerset \to
  \integerset$ such that, for any coloured structure
  $(S,s^\mathcal{V},s^\mathcal{E})$ such that
  $|S|>\eta (N,\mathrm{td}(S))$ and any mapping
  $g:|S|\to\{1,2,\ldots,N\}$, there exists a non-trivial involuting
  $g$-preserving automorphism $\mu$ of
  $(S,s^\mathcal{V},s^\mathcal{E})$ with the fixed-point property. 
\end{thm}
\proof
Let $(S,s^\mathcal{V},s^\mathcal{E})$ be a coloured structure
with $p\geq 1$ connected components
$(S_i,s_i^\mathcal{V},s_i^\mathcal{E})$ where $1\leq i  \leq p$ 
and let $g:|S|\to\{1,2,\ldots,N\}$. Let $F$ be a rooted forest  that
consists of rooted trees $(r_i,Y_i)$, where $1\leq i  \leq p$, such that 
$(r_i,Y_i)$ is an elimination tree for $S_i$ of height at most
$\mathrm{td}(S)$. We prove this result by induction on
$\mathrm{td}(S)$. 

If $\mathrm{td}(S)=1$ then every $(r_i,Y_i)$ has height $1$ and every
$(S_i,s_i^\mathcal{V},s_i^\mathcal{E})$ is a coloured self-loop (a
self-loop is a structure with a single element). Let $|\sigma|$ be the
number of relation symbols in $\sigma$. 
There are at most $N\times|\mathcal{V}|\times |\mathcal{E}|
\times 2^{|\sigma|}$ $g$-valued coloured self-loops. Hence, we set
$\eta (N,1):= N\times|\mathcal{V}|\times |\mathcal{E}| \times
2^{|\sigma|}$ and for any $(S,s^\mathcal{V},s^\mathcal{E})$ satisfying $|S|\geq
\eta (N,1)$ and any mapping $g$ from $|S|$ to $\{ 1,2\ldots,N \}$, there exists a
non-trivial involuting $g$-preserving automorphism $\mu$ of
$(S,s^\mathcal{V},s^\mathcal{E})$ that has the fixed point property
(simply choose for $\mu$ an automorphism that permutes two identical
coloured self-loops with the same $g$ value). In the rest of the
proof, we shall write that $\mu$ is a ``good'' $g$-preserving 
automorphism for short. 

Assume that the result holds for every coloured structure of tree
depth at most $n$, and let $S$ be a structure with tree-depth $n+1$.
Assume first that $p=1$. That is, $F$ consists of a single
rooted tree $(r,Y)$. Let $(r_j,Y_j)$, $1\leq j \leq q$ be the
subtrees of $Y$ where $r_j$ is a child of $r$ in $Y$.  We wish to
define a mapping $g'$ over $S\setminus\{r\}$ such that $g'(x)$
describes both $g(x)$ and the relationship of $x$ and $r$ in
$S$. Let $N(x,r)$ be the coloured structure (together with the
restriction of $g$) that is the substructure of $S$ induced by the
following set of elements: 
$$\bigcup\{e \text{ hyperedge of } \mathcal{H}_S \text{ s.t. }
x\in e \text{ and } r\in e \}.$$
If this set is empty, we set $N(x,r)$ to be $\emptyset$.  Let
$\Theta$ be the set of such neighbourhoods $N(x,r)$ with two constants
$x$ and $r$, considered up to isomorphisms. We also assume that
$\emptyset$ is an element of $\Theta$. Since $\sigma$,
$\mathcal{E}$ and $\mathcal{V}$ are fixed, it follows that $\Theta$ is
finite. We define $g'$ as the mapping from $|S|\setminus\{r\}$ to
$N\times\Theta$ such that $g'(x):=(g(x), \tilde{N}(x,r))$, where
$\tilde{N}(x,r)$ belongs to $\Theta$ and is isomorphic to $N(x,r)$.
Hence, by induction if $|S\setminus \{r\}|>\eta (N|\Theta|,n)$, there
exists a ``good'' $g'$-preserving automorphism $\mu'$ of
$S\setminus\{r\}$. Let $\mu'$ be the extension of $\mu'$ to $S$ such
that $\mu(r):=r$. By construction, $\mu$ is a ``good'' $g$-preserving
automorphism.

Assume now that $p>1$. If one of the components
$(S_i,s_i^\mathcal{V},s_i^\mathcal{E})$ with elimination tree
$(r_i,Y_i)$ has size strictly greater than $\eta(N|\Theta|,n)+1$, 
then from the previous case, it has a ``good'' $g$-preserving
automorphism $\mu_i$ which we can extend by the identity into a
``good'' $g$-preserving automorphism $\mu$ of the whole structure.
Assume now that every component
$(S_i,s_i^\mathcal{V},s_i^\mathcal{E})$ has size at most
$\eta(N|\Theta|,n)+1$. 
The number $\zeta(N,k)$ of coloured structures (together with a
mapping $g$ to a set of size $N$) of size at most $k$ considered up
to isomorphism depends only on $N$ and $k$ (and the constants
$\sigma,\mathcal{V}$ and $\mathcal{E}$).
Hence, if $p>\zeta\bigl(N,\eta(N(|\Theta|,n)+1\bigr)$
then there must be at least two components that are
isomorphic (also w.r.t. both colours and $g$-values).
Hence, $\mu$ is a ``good'' $g$-preserving automorphism, where
$\mu$ is the automorphism that exchanges these two components and leaves the
element of any other component fixed. Thus, we may set, 
$$\eta (N,n+1):= \bigl(\eta(N|\Theta|,n) +1 \bigr)  \times
\zeta\bigl( N,\eta (N|\Theta|,n) +1 \bigr).$$
This concludes the proof.
\qed

\begin{thm}\label{theo:retract}
  Let $(S,s^\mathcal{V},s^\mathcal{E})$ be a coloured structure.
  If $|S|>\eta (1,\mathrm{td}(S))$ then $(S,s^\mathcal{V},s^\mathcal{E})$
  maps homomorphically into one of its proper induced substructure  
  $(S_0,s_0^\mathcal{V},s_0^\mathcal{E})$.
  Consequently, the core of $(S,s^\mathcal{V},s^\mathcal{E})$ has size
  at most $\eta (1,\mathrm{td}(S))$.
\end{thm}
\proof
  By the previous theorem, if $|S|>\eta (1,\mathrm{td}(S))$ then there
  exists a non-trivial involuting automorphism $\mu$ of
  $(S,s^\mathcal{V},s^\mathcal{E})$ with the fixed-point property.
  Let $F$ be the set of fixed-points of $\mu$. Let $S'$ be the
  substructure of $S$ induced by $|S|\setminus|F|$. Let $S''$ be a
  connected substructure of $S'$. Since $\mu$ has the fixed-point
  property and $S''$ has no fixed point by definition, we know that 
  $\mu(S'')\cap S''=\emptyset$. Note that there can not be any
  tuple $t$ in a relation of $S'$ that involves simultaneously 
  elements of $S''$ and $\mu(S'')$ (Otherwise, the substructure
  $S''_{t}$ of $S'$ induced by $|S''|$ and the element of $t$ would be
  connected and $\mu(S''_{t})\cap S''_{t}\neq \emptyset$). There could
  be however a tuple that involves simultaneously elements of
  $F$, $S''$ and $\mu(S'')$. We need a slightly stronger result than
  the previous theorem. Note that in the previous proof, given a
  structure $S$, we can fix a tree decomposition $(r,Y)$ of height
  $\mathrm{td}(S)$ with subtrees $(r_i,Y_i)$ that corresponds to
  substructures $S_i$ of $S$, and subsequently at the induction stage,
  rather than using any tree-decomposition of the structures $S_i$, we
  can assume that $(r_i,Y_i)$ is used instead. With this further assumption,
  we ensure that the automorphism $\mu$ we build has still the
  required properties, and moreover that there is no tuple that
  involves simultaneously elements of $F,S''$ and $\mu(S'')$, for any
  connected substructure $S''$ of $S'$.
  
  Hence, starting with two empty sets $A$ and $B$, we can inductively pick a
  connected component $S''$ of $S'$, add $S''$ to $A$ and $\mu(S'')$
  to $B$ until $S'$ is partitioned into $A$ and $B$. By construction,
  there is no tuple involving elements of both $A$ and $B$ (and possibly
  some element of $F$). Let $h$ be the mapping such that
  $h(x):=\mu(x)$, if $x\in B$, and $h(x):=x$, otherwise. By
  construction of $\mu, A$ and $B$, it follows that $h$ is a
  homomorphism from $(S,s^\mathcal{V},s^\mathcal{E})$ to
  $(S_0,s_0^\mathcal{V},s_0^\mathcal{E})$, the substructure of
  $(S,s^\mathcal{V},s^\mathcal{E})$ induced by $|F|\cup A$. Since
  $\mu$ is non-trivial, $B\neq \emptyset$ and
  $(S_0,s_0^\mathcal{V},s_0^\mathcal{E})$ is a proper induced
  substructure of $(S,s^\mathcal{V},s^\mathcal{E})$.
  This proves the first claim.

  For the second claim, we apply inductively the first claim until we
  get a proper induced substructure
  $(S_\star,s_\star^\mathcal{V},s_\star^\mathcal{E})$  of size at most 
  $\eta (1,\mathrm{td}(S))$. Thus, $(S_\star,s_\star^\mathcal{V},s_\star^\mathcal{E})$
  is an induced substructure of $(S,s^\mathcal{V},s^\mathcal{E})$ in which
  $(S,s^\mathcal{V},s^\mathcal{E})$ maps homomorphically into
  (composing the homomorphisms, we get a homomorphism). The core of
  $(S_\star,s_\star^\mathcal{V},s_\star^\mathcal{E})$ is also the core of
  $(S,s^\mathcal{V},s^\mathcal{E})$ (since the two structures are
  homomorphically equivalent) and its size is at most that of
  $(S_\star,s_\star^\mathcal{V},s_\star^\mathcal{E})$, which is bounded by
  $\eta (1,\mathrm{td}(S))$.
\qed

Thus, we get the following result.
\begin{cor}
  Let $\mathcal{K}$ be any class of coloured structures of bounded tree-depth
  $k$. Then the set $\mathcal{K}'$ of cores of structures from
  $\mathcal{K}$ (up to isomorphism) is finite. \qed
\end{cor}

\subsection{Decompositions.}
\label{sec:decompositions}
Before we can define this concept for structures, let us
briefly recall how the notion was first defined for graphs.
In~\cite{LowTreeWidth}, De Vos~\etal{} proved that for any proper minor
closed class $\mathcal{K}$ and any integer $p\geq 1$, there exists an
integer $q$ such that for every graph $G$ in $\mathcal{K}$ there
exists  a vertex partition of $G$ into $q$ parts such that any
subgraph of $G$ induced  by at most $p$ parts has tree-width at
most $p-1$ (existence of a low tree-width decomposition). Similarly,
we say speak of a \emph{low tree-depth decomposition} for
$\mathcal{K}$ whenever, for every integer $p$, there exists  an
integer $q$ such that any graph in $\mathcal{K}$ has \emph{a proper
$q$-colouring} in which any $p$ colours induce a subgraph of tree-depth at
most $p$.
 
In~\cite{ProperMinorClosedClass}, \Nesetril{} and Ossona de Mendez
refined De Vos~\etal{}'s result to low tree-depth decomposition.
More recently, they reproved this result in the more general setting
of graphs of bounded expansion without using De Vos~\etal{}'s result
and obtained the following characterisation of low tree-depth
decomposition for graphs. 
\begin{thm}{\cite{NesetrildeMendezBoundedExpansion3}} 
  \label{theor:decomposition}
  Let $\mathcal{K}$ be a class of graphs. $\mathcal{K}$ has bounded
  expansion if, and only if, $\mathcal{K}$ has low tree-depth
  decomposition.
\end{thm}

We say that a class $\mathcal{K}$ of structures has \emph{low tree-depth
decomposition} if, and only if, for every $p\geq 1$, there exists an
integer $q$ such that for any structure $S$ in $\mathcal{K}$, there
exists a partition of $|S|$ into $q$ sets such that any substructure
of $S$ induced by at most $p$ of these sets has tree-depth at most $p$.
\begin{prop}\label{cor:decomposition}
  Let $\mathcal{K}$ be a class of structures. If $\mathcal{K}$ has bounded
  expansion then $\mathcal{K}$ has low tree-depth decomposition.
\end{prop}
\proof
  Let $S$ be a structure in $\mathcal{K}$. We apply
  Theorem~\ref{theor:decomposition} to $\mathcal{G}_S$, the Gaifman
  graph of $S$  and get a vertex partition such that every 
  subgraph  of $\mathcal{G}_S$ induced by at most $p$ colours has a
  tree-decomposition of height at most $p$. We use the same partition
  for $S$. For every substructure $S'$ of $S$ induced by at most $p$
  parts, the tree-decomposition of $\mathcal{G}_{S'}$ is also a
  tree-decomposition of $S'$ (by
  Lemma~\ref{lemma:gaifmangraph:structure}) and the result follows.
\qed
\begin{rem}
  We adopt a definition of low tree-depth decomposition of
  structures that seems more natural than a definition that would
  involve the Gaifman graph (we require only a vertex partition). We
  do not know whether our definition is more general or not (in other
  words, does the converse implication in the previous result hold?).
\end{rem}

\subsection{Truncated Product}
We extend the definition of truncated product and adapt two lemmas
from~\cite{ProperMinorClosedClass} to coloured
structures. Recall first that the usual notion of product in the
context of graph homomorphism is the following. The product of two
structures $A$ and $B$ is the structure with universe the cartesian
product of the universes of $A$ and $B$ and such that
$R((a_1,b_1),(a_2,b_2),\ldots,(a_n,b_n))$ holds if, and only if, both
$R(a_1,a_2,\ldots,a_n)$ holds in $A$ and $R(b_1,b_2,\ldots,b_n)$ holds
in $B$. The truncated product resembles this ``classical'' product except
that each component has  a special ``don't care'' element and the
domain of this special product consists only of tuples involving
exactly one ``don't care'' element. 

Let
$(S,s^\mathcal{V},s^\mathcal{E})$ be a coloured structure and 
$p\geq 2$ be an integer. We define the $p$th \emph{truncated product}
of $(S,s^\mathcal{V},s^\mathcal{E})$, to be the coloured structure
$(S',{s'}^\mathcal{V},{s'}^\mathcal{E})$ defined as follows.
\begin{enumerate}[$\bullet$]
\item Its domain is a subset of $\bigcup_{i=1}^{p} W^i$ where,
  $$W^i:=\{(x_1,x_2,\ldots,x_{i-1},\star,x_{i+1},\ldots,x_p)\text{
    s.t. } \forall 1\leq k \leq p, k \neq i \implies x_k \in |S|\}$$
  ($\star$ denotes a new element, \ie{} $\star \not\in |S|$).
\item We restrict further $W^i$ to $\tilde{W}^i$ that consists of
  elements 
  $$w^i = (x_1,x_2,\ldots,x_{i-1},\star,x_{i+1},\ldots,x_p)$$
  of $W^i$ such that there  exists $v\in\mathcal{V}$ such that for
  every $1\leq k \leq p$ with $k\neq i$ we have $s^\mathcal{V}(x_k)=v$
  and we set $|S'|:=\cup_{i=1}^p \tilde{W}^i$ and ${s'}^\mathcal{V}(w^i):=v$.
\item For every relation symbol $R$ of arity $r$, and every tuple
  $(w^{i_1},w^{i_2},\ldots,w^{i_r})$ where for every $1\leq k \leq r$,
  $w^{i_k}$ belongs to $\tilde{W}^{i_k}$ and
  $$w^{i_k}=(x_1^{i_k},x_2^{i_k},\ldots,x_{i_k-1}^{i_k},\star,x_{i_k+1}^{i_k},\ldots,x_p^{i_k})$$
  that satisfies for every $1\leq i \leq p$, with $i\not\in
  \{i_1,i_2,\ldots,i_r\}$:
  \begin{enumerate}[$-$]
  \item $R(t_i)$ holds in $S$, where
    $t_i=(x_i^{i_1},x_i^{i_2},\ldots,x_i^{i_r})$; and, 
  \item there exists $e\in \mathcal{E}$ such that for every $1\leq i \leq p$,
    $s^\mathcal{E}(t_i)=e$,
  \end{enumerate}
  we set $R(w^{i_1},w^{i_2},\ldots,w^{i_r})$ to hold in 
  $S'$ and we set ${s'}^\mathcal{E}(w^{i_1},w^{i_2},\ldots,w^{i_r}):=e$.
\end{enumerate}
We denote the $p$th truncated product by
$(S,s^\mathcal{V},s^\mathcal{E})^{\Uparrow p}$.

This product has two important properties: it preserves validity
\mbox{w.r.t.} a forbidden patterns problem (for a suitable $p$); and, the
existence of all ``partial'' colour-preserving homomorphisms is equivalent
to the existence of a homomorphism to the truncated product (see
Lemma~\ref{lem:truncated-product:preserve:validity} and
Lemma~\ref{lem:partial:homs:to:hom:base:case} below).
\begin{lem}\label{lem:truncated-product:preserve:validity}
  Let $p\geq 2$.
  Let $\mathfrak{F}$ be a set of forbidden patterns such that for every
  $(F,f^\mathcal{V},f^\mathcal{E})$ in $\mathfrak{F}$, we have  $|F|<
  p$. If a coloured structure $(S,s^\mathcal{V},s^\mathcal{E})$ is
  valid w.r.t $\mathfrak{F}$ then its $p$-truncated product
  $(S',{s'}^\mathcal{V},{s'}^\mathcal{E})$ is also valid w.r.t. $\mathfrak{F}$.  
\end{lem}
\proof
We prove the contrapositive.
Assume that $(S',{s'}^\mathcal{V},{s'}^\mathcal{E})$ is not valid
w.r.t. $\mathfrak{F}$, that is for some
$(F,f^\mathcal{V},f^\mathcal{E})$ in $\mathfrak{F}$, there exist a
colour-preserving homomorphism $f'$ from $F$ to $S'$.
Since $|F|< p$, by the pigeon-hole principle there exists $1\leq i_0
\leq p$ such that $f'(F)\cap \tilde{W}^{i_0} = \emptyset$. 

Let $\pi_{i_0}$ be the mapping from $|S'|\setminus \tilde{W}^{i_0}$ to
$|S|$ defined as $\pi_{i_0}(w^i)=x^i_{i_0}$ (with the same notation as
above). It follows directly from the definition of $S'$ that
$\pi_{i_0}$ is a homomorphism from $S'\setminus \tilde{W}^{i_0}$ to
$S$, where $S'\setminus \tilde{W}^{i_0}$ is the substructure of $S'$
induced by $|S'|\setminus \tilde{W}^{i_0}$. 

Hence, by composition $f:=\pi_{i_0}\circ f'$ is a colour-preserving
homomorphism from $F$ to $S$ and we have proved that
$(S,s^\mathcal{V},s^\mathcal{E})$ is not valid w.r.t. $\mathfrak{F}$.
\qed

\begin{lem}\label{lem:partial:homs:to:hom:base:case}
  Let $(U,u^\mathcal{V},u^\mathcal{E})$ be a coloured structure and let
  $p$ be an integer greater than the  arity of any symbol in $\sigma$.
  Let $(S,s^\mathcal{V},s^\mathcal{E})$ be a coloured structure.
  If there exists a partition $V_1,V_2,\ldots,V_p$ of $|S|$ such that
  for every substructure
  $(\tilde{S}_i,\tilde{s}^\mathcal{V},\tilde{s}^\mathcal{E})$ of
  $(S,s^\mathcal{V},s^\mathcal{E})$  induced by $|S|\setminus V_i$
  there exist a colour-preserving homomorphism $\tilde{s}_i$  from
  $(\tilde{S}_i,\tilde{s}^\mathcal{V},\tilde{s}^\mathcal{E})$ to
  $(U,u^\mathcal{V},u^\mathcal{E})$  then there exists a
  colour-preserving homomorphism $\tilde{s}$ from
  $(\tilde{S}_i,\tilde{s}^\mathcal{V},\tilde{s}^\mathcal{E})$ to
  $(U,u^\mathcal{V},u^\mathcal{E})^{\Uparrow p}$.
\end{lem}
\proof
  Let $(U',{u'}^\mathcal{V},{u'}^\mathcal{E}):=
  (U,u^\mathcal{V},u^\mathcal{E})^{\Uparrow p}$.  
  Let $x$ be an element of $S$ such that $x$ belongs to $V_i$.
  Then $x$ is an element of $\tilde{S}_k$ for every $1\leq k \leq p$
  such that $k\neq i$ and $\tilde{s}_k(x)$ is defined and we set:
  $$\tilde{s}(x):=(\tilde{s}_1(x),\tilde{s}_2(x),\ldots,\tilde{s}_{i-1}(x),
  \star,\tilde{s}_{i+1}(x),\ldots,\tilde{s}_{p}(x)).$$
  Since $\tilde{s}_k$ is colour-preserving, 
  $u^\mathcal{V}(\tilde{s}_k(x))= \tilde{s}^\mathcal{V}(x)$ and
  $\tilde{s}(x)$ is indeed in $|U'|$ by the definition of
  the truncated product.
  
  Let $x_1,x_2,\ldots,x_r$ be elements of $S$ that belong to the sets
  $V_{i_1},V_{i_2},\ldots,V_{i_r}$, respectively. Let $R$ be a $r$-ary
  relation symbol from $\sigma$ such that $R(x_1,x_2,\ldots,x_r)$
  holds in $S$. Then, for every $1\leq k \leq p$ such that
  $k\not\in\{i_1,i_2,\ldots,i_r\}$, we  have that
  $R(\tilde{s}_k(x_1),\tilde{s}_k(x_2),\ldots,\tilde{s}_r(x_r))$ holds
  in $U$. Moreover, since $\tilde{s}_k$ is colour-preserving, we have
  that
  $u^\mathcal{E}(\tilde{s}_k(x_1),\tilde{s}_k(x_2),\ldots,\tilde{s}_r(x_r)) 
  =\tilde{s}^\mathcal{E}(x_1,x_2,\ldots,x_r)$
  and it follows by the
  definition of the truncated product that\\
  $R(\tilde{s}(x_1),\tilde{s}(x_2),\ldots,\tilde{s}(x_r))$ holds in
  $U'$ and that $\tilde{s}$ is a colour-preserving homomorphism.
\qed

Using the two previous lemmas, an easy induction provides the following result.
\begin{prop}\label{prop:induction:step}
  Let $p$ be an integer greater than the arity of any symbol in $\sigma$.
  Let $\mathfrak{F}$ be a set of forbidden patterns such that, for every
  $(F,f^\mathcal{V},f^\mathcal{E})$ in $\mathfrak{F}$, we have  $|F|<
  p$.
  Let $q\geq p$.
  Let $(U',{u'}^\mathcal{V},{u'}^\mathcal{E})$ be a coloured structure that is
  valid w.r.t $\mathfrak{F}$. Let $(S,s^\mathcal{V},s^\mathcal{E})$
  be a coloured structure.  

  Assume that there exists a partition $V_1,V_2,\ldots,V_q$ of $|S|$, such that
  for every substructure
  $(\tilde{S}_i,\tilde{s}_i^\mathcal{V},\tilde{s}_i^\mathcal{E})$ of
  $(S,s^\mathcal{V},s^\mathcal{E})$  induced by $p$ subsets
  $V_{i_1},V_{i_2},\ldots,V_{i_p}$,  
  there exists a colour-preserving homomorphism $\tilde{s}_i$  from
  $(\tilde{S}_i,\tilde{s}_i^\mathcal{V},\tilde{s}_i^\mathcal{E})$ to
  $(U',{u'}^\mathcal{V},{u'}^\mathcal{E})$.

  Then there exists a colour-preserving homomorphism $\tilde{s}$ from
  $(\tilde{S}_i,\tilde{s}^\mathcal{V},\tilde{s}^\mathcal{E})$ to
  $(U,u^\mathcal{V},u^\mathcal{E})$ and
  $(U,u^\mathcal{V},u^\mathcal{E})$ is valid with respect to
  $\mathfrak{F}$, where 
  $$(U,u^\mathcal{V},u^\mathcal{E}):=
  (U',{u'}^\mathcal{V},{u'}^\mathcal{E})^{\Uparrow (p+1)\Uparrow (p+2) \ldots
  \Uparrow  q}.$$\qed
\end{prop}

\subsection{Universal Structure}
We can now conclude the proof of
Theorem~\ref{thm:coloured:restricted:duality:bounded:expansion}, whose
statement we recall now.
\begin{thm*}
  Let $\mathcal{C}$ be a class of structures. 
  If $\mathcal{C}$ has low tree-depth decomposition (\eg{} bounded
  degree, proper minor closed class, structure of bounded expansion)
  then $\mathcal{C}$ has all restricted coloured dualities. 
\end{thm*}
\proof
  Let $(S,s^\mathcal{V},s^\mathcal{E})$ be a coloured structure such
  that $S$  belongs  to $\mathcal{K}$. 
  By Corollary~\ref{cor:decomposition}, there exists an integer $N$
  such that every structure $S$ in $\mathcal{K}$ can be partitioned
  into $q$ parts,  such that every $p$ parts induce a substructure of
  $S$ of tree-depth at most $p$. By Theorem~\ref{theo:retract}, the
  core of the coloured structure induced by $p$ parts is bounded. 
  Let $(U',{u'}^\mathcal{V},{u'}^\mathcal{E})$ be the disjoint union of all
  such cores that are valid w.r.t. $\mathfrak{F}$  (there  are only
  finitely many). Since the forbidden patterns
  have size at most $p$, it suffices to check every substructure of
  $S$ of size at most $p$ and, \emph{a fortiori},
  $(S,s^\mathcal{V},s^\mathcal{E})$ is valid w.r.t. $\mathfrak{F}$ if,
  and only if, each of its coloured substructures induced by $p$ parts
  is valid, or equivalently  if every such coloured substructure maps
  homomorphically into $(U',{u'}^\mathcal{V},{u'}^\mathcal{E})$.

  Let $(U,u^\mathcal{V},u^\mathcal{E}):=
  (U',{u'}^\mathcal{V},{u'}^\mathcal{E})^{\Uparrow (p+1)\Uparrow (p+2)
    \ldots \Uparrow  q}$.
  By Proposition~\ref{prop:induction:step},
  if $(S,s^\mathcal{V},s^\mathcal{E})$ is  valid w.r.t. $\mathfrak{F}$
  then it is homomorphic to 
  $(U,u^\mathcal{V},u^\mathcal{E})$; and, since
  $(U,u^\mathcal{V},u^\mathcal{E})$ is valid w.r.t. $\mathfrak{F}$ the
  converse holds  as forbidden patterns problems are closed under
  inverse homomorphism. 
  
  As in the proof of
  Theorem~\ref{thm:coloured:restricted:duality:bounded:degree} in the 
  previous section, forgetting the colours provides us with the desired
  template $U$.
\qed
\section{Logical aspects}
\label{sec:logical-aspects}

\subsection{MMSNP$_1$ and MMSNP$_2$}
The class of forbidden patterns problems with colours over vertices only,
corresponds to the problems that can be expressed by a formula in Feder and
Vardi's MMSNP (Monotone Monadic SNP without inequalities,
see~\cite{FederVardi,ThePaperLong}). Note that allowing 
colours over the edges does not amount to  dropping the hypothesis of
monadicity altogether. Rather, it corresponds to a logic, let's
call it MMSNP$_2$, which is similar to MMSNP but allows
\emph{first-order} variables over edges (just like Courcelle's MSO
(Monadic Second Order logic) and MSO$_2$,
see~\cite{CourcelleMSO6}). 

In the following, we introduce formally MMSNP$_1$ and recall some
known results; and, secondly, we introduce MMSNP$_2$ and prove that
it defines finite unions of problems in FPP$_2$. 

\begin{defi}\label{defMMSNP}
  \emph{Monotone Monadic SNP without inequality}, MMSNP$_1$, is the
  fragment of MSO consisting of those formulae
  $\Phi$ of the following form:
  $$\exists \mathbf{M} \forall \mathbf{t} \bigwedge_i
  \lnot\bigl(\alpha_i(\sigma,\mathbf{t}) \land
  \beta_i(\mathbf{M},\mathbf{t})\bigr),$$  
  where $\mathbf{M}$ is a tuple of monadic relation symbols (not in
  $\sigma$), $\mathbf{t}$ is a tuple of (first-order) variables and
  for every negated conjunct $\lnot(\alpha_i \land \beta_i)$:
  
  \begin{enumerate}[$\bullet$]
  \item $\alpha_i$ consists of a conjunction of positive atoms
    involving relation symbols from $\sigma$ and variables from
    $\mathbf{t}$; and  
  \item $\beta_i$ consists of a conjunction of atoms or negated
    atoms involving relation symbols from $\mathbf{M}$ and variables
    from $\mathbf{t}$.
  \end{enumerate}
  
  \noindent (Notice that \emph{the equality symbol does not occur\/}
  in $\Phi$.)
\end{defi}
The negated conjuncts $\lnot (\alpha \land \beta)$ correspond to
(partially coloured) forbidden structures (and this is the reason why
we use such a notation in the definition rather than using
implications or clausal form). To get forbidden patterns
problems, we need to restrict sentences so that negated conjuncts
correspond precisely to coloured connected structures. Such a
restriction was introduced in~\cite{ThePaperLong} as follows.
\begin{defi}
  \label{primitive}
  Let $\Phi$ be as in Definition~\vref{defMMSNP}.
  $\Phi$ is \emph{primitive\/} if, and only if, moreover,
  for every negated conjunct $\lnot(\alpha \land \beta)$:
  
  \begin{enumerate}[$\bullet$]
  \item for every first-order variable $x$ that occurs in
    $\lnot(\alpha \land \beta)$ and for every monadic symbol $C$ in
    $\mathbf{M}$, exactly one of $C(x)$ and $\lnot C(x)$ occurs in
    $\beta$;
    
  \item unless $x$ is the only first-order variable that occurs in
    $\lnot(\alpha \land \beta)$, an atom of the form $R(\mathbf{t})$,
    where $x$ occurs in $\mathbf{t}$ and $R$ is a relation symbol from
    $\sigma$, must occur in $\alpha$; and,

  \item the structure induced by $\alpha$ is connected. 
  \end{enumerate}
\end{defi}
\begin{rem}
  We have altered slightly the definitions
  w.r.t.~\cite{ThePaperLong}. We now require a pattern to be
  connected. However, we have amended the notion of a primitive
  sentence accordingly. Thus, the following statement still holds as
  the connectivity requirement is enforced on both sides of the
  equivalence.
\end{rem}
\begin{thm}{\cite{ThePaperLong}}\label{theo:MMSNP:captures:FPP}
  The class of problems defined by the primitive fragment of the
  logic MMSNP$_1$ is exactly the class FPP$_1$ of forbidden
  patterns problems with vertex colours only.
\end{thm}
It is only a technical exercise to relate any sentence of MMSNP$_1$
with its primitive fragment.
\begin{prop}{\cite{ThePaperLong}}\label{pro:collection:of:primitive}
  Every sentence of {\bf MMSNP$_1$} is logically equivalent to a finite
  disjunction of primitive sentences.
\end{prop}
This paper is concerned with decision problems only and we equate a
problem with the (isomorphism closed) set of its yes-instances. Thus,
we may speak of the \emph{union of two problems}.
Consequently, we have the following characterisation.
\begin{cor}\label{cor:MMSNP1vsFPP1}
  Every sentence $\Phi$ in MMSNP$_1$ defines the union of finitely
  many problems in  FPP$_1$. 
\end{cor}

The logic MMSNP$_2$ is the extension of the logic MMSNP$_1$ where in
each negated conjunct $\lnot(\alpha\land\beta)$, we allow a monadic 
predicate to range over a tuple of elements of the structure, that is
we allow new ``literals'' of  the form $M(R(x_1,x_2,\ldots,x_n))$ in $\beta$,
where $R$ is $n$-ary relation symbol from the signature
$\sigma$. We also insists that whenever such a literal occurs in
$\beta$ then $R(x_1,x_2,\ldots,x_n)$ appears in $\alpha$. 
The semantic of a monadic predicate $M$ is extended and is defined as
both a subset of the domain and a subset of the set of tuples that
occur in some input 
relation: that is, for a structure $S$, $M^S\subset |S| \cup E(S)$. We
say that a sentence of MMSNP$_2$ is 
\emph{primitive} if each negated conjunct $\lnot(\alpha\land \beta)$
satisfies the same conditions as in Definition~\ref{primitive} and a
further condition:
\begin{enumerate}[$\bullet$]
\item if $R(x_1,x_2,\ldots,x_n)$ occurs in $\alpha$ then for every
  (existentially quantified) monadic predicate $C$ exactly one of 
  $C(R(x_1,x_2,\ldots,x_n))$ or $\lnot C(R(x_1,x_2,\ldots,x_n))$ occurs
  in $\beta$.
\end{enumerate}

It is only a technical exercise to extend all the previous results of
this section concerned with MMSNP$_1$ and FPP$_1$ to MMSNP$_2$  and
FPP$_2$. In particular, we have the following result.
\begin{cor}\label{cor:MMSNP2vsFPP2}
  Every sentence $\Phi$ in MMSNP$_2$ defines
  the union of finitely many problems in FPP$_2$.\qed
\end{cor}

\subsection{Edge Quantification versus Vertex Quantification}
Courcelle investigated the difference in expressivity that adding
edge set quantification provided to MSO: he proved that
MSO$_2$ (with edge set quantification) is more expressive than MSO$_1$
(with the more usual vertex set quantification) in general. 
However, he also showed that under certain restriction, edge set
quantification does not add to MSO's expressivity.   
\begin{thm}{\cite{CourcelleMSO6}}
  \label{thm:Courcelle:collapse:old}
  On each of the following classes of simple graphs: 
  those of degree at most $k$, those of tree-width at most $k$, for
  each $k$, planar graphs, and, more generally, every \emph{proper
    minor closed class}, every sentence in MSO$_2$ is logically
  equivalent to a sentence of MSO$_1$. 
\end{thm}

Restricted coloured duality theorems can be reformulated in terms of
expressivity of MMSNP$_2$, MMSNP$_1$ and constraint satisfaction problems as the following result
shows. 
\begin{thm}{(\textbf{Collapse to CSP}).}
  \label{thm:coloured:restricted:duality:reformulator}
  If a class $\mathcal{K}$ has all restricted coloured dualities then 
  MMSNP$_1$ and MMSNP$_2$ are equally expressive when restricted to
  inputs from $\mathcal{K}$. These logics define precisely finite unions of
  constraint satisfaction problems; and, in particular if
  $\mathcal{K}$ contains connected structures only then these logics
  define precisely constraint satisfaction problems. 
\end{thm}
\proof
  By Corollary~\ref{cor:MMSNP1vsFPP1} (resp.
  Corollary~\ref{cor:MMSNP2vsFPP2}) every problem in MMSNP$_1$ (resp.
  MMSNP$_2$) is a finite  union of problems from FPP$_1$
  (resp. FPP$_2$). When restricted to a class $\mathcal{K}$ that has
  all restricted coloured dualities, every forbidden patterns
  problem is a restricted CSP. Moreover, any finite union of constraint satisfaction problems
  can be written using a sentence in
  MMSNP$_1$~\cite{FederVardi,ThePaperLong}.
  This proves that the logics MMSNP$_1$ and
  MMSNP$_2$, when restricted to a class $\mathcal{K}$ that has all
  restricted coloured dualities, define precisely finite unions of
  constraint satsifaction problems.
  Moreover, if we assume that the input is also connected, then a
  finite union of constraint satisfaction problems is a single
  constraint satisfaction problem: indeed, its template consists of 
  the disjoint union of the template  of each constraint satisfaction
  problem. 
\qed
\begin{rem}
  An alternative proof strategy would be to use Courcelle's method
  from~\cite{CourcelleMSO6} to build a sentence of MSO$_1$ and next to
  transform it hopefully into an equivalent sentence of MMSNP$_1$ using
  preservation under inverse homomorphisms as in Feder and Vardi's work
  on preservation theorem for MMSNP in~\cite{FederVardi03}. 
  However, even if successful such a proof strategy would only provide
  a proof that both logics become equally expressive and would not
  provide the collapse to (finite union of) CSP(s) as required.
\end{rem}
Thus, Theorem~\ref{thm:coloured:restricted:duality:bounded:degree} and 
Theorem~\ref{thm:coloured:restricted:duality:pmcc} reformulated using
Theorem~\ref{thm:coloured:restricted:duality:reformulator} provides us
with the analogous result for MMSNP to Courcelle's result for MSO.
\begin{cor}
  \label{cor:MMSNP:collapse:a:la:courcelle}
  On each of the following classes of simple graphs: 
  those of degree at most $k$, those of tree-width at most $k$, for
  each $k$, planar graphs, and, more generally, every \emph{proper
    minor closed class}, 
  every sentence in MMSNP$_2$ is logically equivalent to a sentence of
  MMSNP$_1$.\qed
\end{cor}
 
Courcelle has recently extended Theorem~\ref{thm:Courcelle:collapse:old}
to hypergraphs, which can be stated as follows in the case of graphs.  
\begin{thm}{\cite{CourcelleMSO14}}
  \label{thm:Courcelle:collapse:new}
  Let $k>0$.
  Every sentence of MSO$_2$ is logically equivalent to a sentence of
  MSO$_1$ over uniformly $k$-sparse graphs. 
\end{thm}
Recall that a graph $G$ is \emph{uniformly $k$-sparse} if, and only if,
every subgraph $H$ of $G$ is $k$-sparse, that is $|E(H)|\leq
k.|V(H)|$. This definition is equivalent to the following condition:
$G$ has an orientation such that every vertex has in-degree at most $k$ (see
Lemma 3.1 in~\cite{CourcelleMSO14}). 
\begin{rem}
  It follows directly from the definitions that a class of graphs with
  bounded expansion is uniformly $k$-sparse for some fixed $k$. 
  However, we know that the converse implication can not hold as
  $2$-sparse graphs do not have all restricted dualities.
  Indeed, we prove in Proposition~\ref{pro:no:duality:2:sparse} (see
  below) that 
  there exists a problem definable by a first-order sentence of MMSNP$_1$ 
  that is not a CSP even when restricted to uniformly $2$-sparse graphs.
  However, this does not exclude that MMSNP$_1$ and MMSNP$_2$ are also
  equally expressive when restricted to uniformly $k$-sparse graphs. 
\end{rem}
The following was observed independently
in~\cite{NesetrilPomBookk03}. We provide our own proof for
completeness.  
\begin{prop}
  \label{pro:no:duality:2:sparse}
  Uniformly $2$-sparse graphs do not have all restricted dualities.
\end{prop}
\proof
Consider the problem \textsc{Tri-Free} whose yes-instances are
triangle-free graphs. Consider the graph $G_n$ with $n$ special
elements, such that every pair of distinct special elements are linked
by a path of length three (using additional vertices). 
We give an orientation to each edge on each of the path as follows:
edges with a special vertex become arcs 
originating from this special vertex; and other edges are oriented
arbitrarily. Note that every special vertex has in-degree zero and every
non-special vertex has in-degree at most 2 (since it has degree at
most 2). This shows that our graph is uniformly $2$-sparse.
 
Moreover, by construction $G_n$ is triangle-free and no homomorphic image
of this graph can identify special elements. The family of graphs
$(G_n)_{n\in\integerset}$ provides us with proofs (so-called witness
family in~\cite{ThePaperLong}) that the problem can not be a finite
union of (finite) constraint satisfaction problems even when restricted to graphs that are
uniformly $2$-sparse.
\qed

\subsection{Infinite constraint satisfaction problems and MMSNP$_2$}
Bodirsky \etal{} have investigated constraint satisfaction problems 
where the template is infinite. They have proposed restrictions that ensure that
the problems are decidable (and in $\mathcal{NP}$): when the template is
countable and \emph{homogeneous} in~\cite{ManuelNesetrilCSL03}, and more
recently to a more general case when the template is
$\omega$-categorical in~\cite{ManuelPhd}. Recall that a countable structure
$\Gamma$ is \emph{$\omega$-categorical} if all countable models of the
first-order theory of $\Gamma$ are isomorphic to $\Gamma$. 
Denote by CSP$^\star$ the set of constraint satisfaction problems that
have a $\omega$-categorical countable template and belong to
$\mathcal{NP}$. 
\begin{rem}
  In~\cite{ManuelPhd}, the definition of CSP$^\star$ is more
  restricted than ours. A template is required to be both $\omega$-categorical
  and \emph{finitely constrained} (see definition below). However, the
  property of being finitely constrained is only used in order to
  enforce that the problem belongs to $\mathcal{NP}$. This motivates 
  our more general definition.
  A countable structure $\Gamma$ is
  \emph{finitely constrained} if there is a first-order expansion
  $\Gamma'$ of $\Gamma$ over some expanded signature $\tau'$ and a
  finite set $\mathcal{N}'$ of finite $\tau'$-structures such that
  $Age(\Gamma')=Forb(\mathcal{N}')$, where $Age(\Gamma)$, the
  so-called \emph{age of $\Gamma$} is the set of all finite induced
  substructures of $\Gamma$; and, by $Forb(\mathcal{N})$, we denote 
  the set of all finite structures that do not admit any of the
  structures from $\mathcal{N}$ as an induced substructure. 
\end{rem}

We say that a problem $\Omega$ is \emph{closed under
disjoint union} if for every structures $A$ and $B$, their disjoint
union $A+B$ is a yes-instance of $\Omega$ whenever both $A$ and $B$
are yes-instances of $\Omega$.
Using a recent result due to Cherlin, Shelah and
Chi~\cite{CherlinShelahShi}, Bodirsky and Dalmau proved the following 
result. 
\begin{thm}{\cite{BodirskyDalmauStacs06}}
  \label{theo:MMSNP1:infiniteCSP}
  Every non-empty problem in MMSNP$_1$ that is closed under disjoint
  union belongs to CSP$^\star$.
\end{thm}
It follows directly from the definition that every problem in FPP$_1$
is closed under disjoint union. Hence, we get the following result. 
\begin{cor}
  Every problem in FPP$_1$ is in CSP$^\star$.
  Consequently, every problem in MMSNP$_1$ is the union of finitely
  many problems in CSP$^\star$.
\end{cor}
Since $\omega$-categoricity is preserved under first-order
interpretation, we can prove the following.
\begin{thm}\label{theo:MMSNP2:infiniteCSP}
  Every problem in FPP$_2$ is in CSP$^\star$.
  Consequently, every problem in MMSNP$_2$ is the union of finitely
  many problems in CSP$^\star$.
\end{thm}
\proof

Note that the second claim follows from the first claim together
with Corollary~\ref{cor:MMSNP2vsFPP2}. We now prove the first
claim in the case of a problem $\Omega$ in FPP$_2$. To simplify the 
notation, we assume that $\Omega$ is a problem over digraphs encoded
using a single binary relation $E$. For simplicity, we may also
assume that there are only arc colours (vertex colours can be easily
encoded using more arc colours and additional forbidden
patterns). Let $\mathcal{E}$ be the set of arc colours and let
$|\mathcal{E}|=c$. 

Consider the problem $\Omega'$ in MMSNP$_1$ defined over a structure
with one monadic predicate $T$ and one ternary predicate $R$, given
by a sentence $\Phi$ with $m = \lceil \log c  \rceil$ monadic
predicates. 
We use these monadic predicates to encode the $c$ colours in a natural
way, using a conjunction $\chi_a$ of $m$ monadic predicates for each
colour $a$ (if necessary, $\Phi$ may contain negated conjuncts
forbidding certain combinations of monadic predicates if $2^m >
c$). Each forbidden pattern $(F,f^\mathcal{E})$ of $\Omega$ is
encoded as a negated conjunct as follows: 
for each arc $E(x,y)$ of colour $a$, add the literals
$R(x,e,y)$, $T(e)$ and $\chi_a(e)$. 

The following formula provides an interpretation $\Pi$ of the
signature $\langle E \rangle$
in the signature $\langle T,R \rangle$:
$\psi_E(x,y)= \exists e T(e) \land R(x,e,y)$. By construction of
$\Omega'$, this interpretation $\Pi$ is a  first-order reduction from
$\Omega'$ to $\Omega$. Note also that $\Omega'$ is closed under
disjunction (since every forbidden pattern of $\Omega$ is connected
by definition of FPP$_2$). By Theorem~\ref{theo:MMSNP1:infiniteCSP},
there exists an $\omega$-categorical structure $\Gamma'$ such that
$\Omega'=CSP(\Gamma')$. Let $\Gamma$ be $\Pi(\Gamma')$. By
Proposition 2.7 of~\cite{ManuelPhd} (page 27), it follows  that
$\Gamma$ is also $\omega$-categorical. Moreover, it is not difficult
to check that $\Omega =  CSP^\star(\Gamma)$. 
\qed

\begin{rem}
  As pointed out in~\cite{ManuelPhd}, there are problems that are in
  CSP$^\star$ but not in MMSNP$_1$. For example, the problem over
  directed graphs with template induced by the linear order over
  $\rationalset$. Unfortunately, this problem is not expressible in
  MMSNP$_2$ either. In fact, we do not know whether MMSNP$_1$ is
  strictly contained in MMSNP$_2$. Indeed, to the best  of our
  knowledge, none of the problems used to separate MSO$_1$ from
  MSO$_2$ are expressible in MMSNP$_2$.  
  We suspect that the problem \textsc{Edge-No-Mono-Tri} is not
  expressible in MMSNP$_1$ and that MMSNP$_1$ is strictly contained in
  MMSNP$_2$.
\end{rem}

\section{Conclusion}

\subsection*{Our results}
In this paper, we have proved that every forbidden patterns (with
colours on both edges and vertices) problem is in fact a constraint
satisfaction problem, when  restricted to a class of structures that
have low tree-depth decomposition: \eg{} bounded degree structures, a
proper minor closed class of structures and more generally a class of
bounded expansion. We derive from this result that the logic MMSNP$_2$
(and MMSNP$_1$) coincides with  the class of constraint satisfaction
problems on connected inputs that belong to a class that has low
tree-depth decomposition.  Together these results cover the
restrictions considered by Courcelle in~\cite{CourcelleMSO6} under 
which MSO$_1$ and MSO$_2$ have the same expressive power. 

\subsection*{Some technical questions}
Note that we do not know whether for unrestricted inputs, MMSNP$_2$ is more
expressive than MMSNP$_1$. Moreover, we have seen that Courcelle's more
recent generalisation to uniformly $k$-sparse
graphs~\cite{CourcelleMSO14} does not have an analog for MMSNP$_1$,  
MMSNP$_2$ and constraint satisfaction problems. By this we mean that
the two logics could well be equally expressive under this
restriction, but they must necessarily capture problems that are not
constraint satisfaction problems even when restricted to uniformly 
$k$-sparse graphs, for any fixed $k\geq 2$.

Another point concerns the notion of a proper minor closed class of
\emph{structures}. In the present paper, we use the Gaifman graph to
define this concept. However, it would be more natural and perhaps
preferable to define a notion of \emph{minor for structures}. The
following definition seems reasonable.
A minor of a structure $S$ is obtained from $S$
by performing a finite sequence of the following operations:
taking a (not necessarily induced) substructure; and,
identifying some elements, provided that they all occur in some
tuple of some relation.
This new definition subsumes the definition used in this paper and
provokes the following question: \emph{Do the results of this paper hold
  under this new definition?} 
A similar question arises for a class of structures with bounded
expansion. In particular, is there a suitable definition
that would be equivalent to low tree-depth decomposition (recall that
to define this notion over structures we do not use the Gaifman
graph)? 

\subsection*{Future work}
Apart from the above, perhaps technical questions, there are two general
questions that we plan to investigate in future work. The first
question is related to CSP$^\star$, the class of (well-behaved)
infinite constraint satisfaction problems introduced by Bodirsky. We
know that any problem in MMSNP$_2$ is a finite union of problems from
CSP$^\star$. However, there are problems in CSP$^\star$ that are not
expressible in MMSNP$_2$, which yields the following
question. \emph{Which logic (necessarily, some extension of
  MMSNP$_2$) defines precisely CSP$^\star$?} 

The second one concerns restricted duality and restricted coloured
duality. We have given some ad-hoc techniques to lift results of
restricted duality to restricted coloured duality for some of our
examples. These techniques do not need to know precisely the
construction of the universal graph in the case of restricted duality
they reduce to. Instead they rely on the rather restricted form of our
examples. We wonder whether this is indeed an artifact of the
restricted nature of our examples or whether this can be done in
general. In other words, \emph{Is it the case that all restricted
  dualities for a class $\mathcal{K}$ implies all restricted coloured
  dualities for $\mathcal{K}$?}

\subsection*{Related work}
Independently to our work, Kun and \Nesetril{} have initiated an 
elegant new approach by means of \emph{lifts} and \emph{shadows}
in~\cite{KunNesetril06}. In the sequel paper~\cite{KunNesetril07},
they state a theorem, which corresponds to our main result, but in the
case of vertex colours only. The approach these authors 
propose relies on the fact that  (restricted) coloured duality can be
reduced to (restricted) duality over an extended signature,
where the additional symbols are monadic and used to encode the vertex
colours. Since the property of being of bounded expansion does not
depends on the monadic predicates, this means that one only needs to
generalise the results concerning restricted duality for bounded expansion due
to \Nesetril{} and Ossona de Mendez to arbitrary relational signature. In
other words, the consideration of vertex colouring in our proof has
become unecessary. We believe that it is possible to deal similarly
with edge colours by defining a suitable notion of lift and follow the
lines of Kun and \Nesetril's approach. This would allow us to remove the
consideration of edge-colouring in our proof, simplifying it further.

\section*{Acknowledgement} The author wishes to thank Manuel Bodirsky,
Bruno Courcelle, Anuj Dawar, Arnaud Durand and Barnaby Martin for
useful discussions and comments. The author is thankful to the first
anonymous referee for advice on how to improve the exposition of this
paper, and to the second referee for pointing out the implications
of the independent work carried out by Kun and \Nesetril.

\bibliographystyle{plain}
\bibliography{RestrictedDualitiesBiblio}
\end{document}